\documentclass[12pt]{article}
\usepackage{geometry}
\usepackage{amsmath,amssymb}
\usepackage[nosort]{cite}
\usepackage{float}
\usepackage{latexsym}
\usepackage{graphicx}
\usepackage{color}
\usepackage{subfig}

\newcommand{\Tr}{{\rm Tr\,}}

\renewcommand{\d}{{\rm d}} 
\newcommand{\beq}{\begin{equation}}
\newcommand{\be}{\begin{equation}}
\newcommand{\ee}{\end{equation}}
\newcommand{\bea}{\begin{eqnarray}}
\newcommand{\eea}{\end{eqnarray}}

\newcommand{\pa}{\partial}

\newcommand{\pah}{\hat{\partial}}

\newcommand{\nn}{\nonumber}

\newcommand{\rh}{\hat{\rho}}
\newcommand{\xh}{\hat{x}}
\newcommand{\lh}{\hat{\lambda}}
\newcommand{\xt}{\tilde{x}}
\newcommand{\vt}{\tilde{V}}
\begin{document}

\begin{titlepage}
\hbox to \hsize{\hspace*{0 cm}\hbox{\tt }\hss
    \hbox{\small{\tt TAUP-2892/09}}}

\vspace{1 cm}

\centerline{\bf \Large Holographic Dual of QCD from  }

\vspace{.6cm}

\centerline{\bf \Large Black D5 Branes. }

\vspace{.6cm}

\centerline{\bf \large }

\vspace{1 cm}
 \centerline{\large $^\dagger\!\!$ Benjamin A. Burrington, \large $\;^{*\!\!}$ Jacob Sonnenschein}

\vspace{0.5cm}
\centerline{\it ${}^\dagger$Department of Physics,}
\centerline{\it University of Toronto,}
\centerline{\it Toronto, Ontario, Canada M5S 1A7. }

\vspace{.5 cm}
\centerline{$^{*}${\it School of Physics and Astronomy,}}
\centerline{\it The Raymond and Beverly Sackler Faculty of Exact Sciences,}
\centerline{\it Tel Aviv University, Ramat Aviv, 69978, Israel.}

\vspace{0.3 cm}

\begin{abstract}
We study the dynamics of probe D7 $\overline{\rm D7}$ flavor probe branes in the background of near extremal  D5 branes.  This model is a holographic dual to a gauge theory with  spontaneous breaking of a  $U(N_f)_L\times U(N_f)_R$ chiral symmetry.  The spectrum of  two such D7 $\overline{\rm D7}$ embeddings, contains  a single massive 4D meson coming from the world volume $U(1)$ gauge field, the pion, and a single massive 4D scalar meson coming from fluctuations of the embedding of the brane.  In addition, there are continuum five dimensional states due to the finite height of the effective potential in the radial direction.  We investigate baryons in this model, and find that the size is stabilized due to the Chern-Simons term in the D7 world volume action. The model admits a  Hagedorn temperature of $\frac{1}{2\pi R}$ where $R$ is the radius  parameter in  the D5 branes metric.
We investigate the pattern of chiral symmetry breaking in the deconfined phase as a function of the asymptotic separation of the branes $L$.  We find that for $\frac{\pi}{3}R<L\lessapprox 1.068 R$ that chiral symmetry is restored, and that chiral symmetry is broken for $L$ outside this window.  We further argue that the solutions with $L<\frac{\pi}{3}$ are only classically stable, and in fact no D7 embedding exists with these boundary conditions.
\end{abstract}
\end{titlepage}

\section{Introduction}

QCD at strong coupling has several properties of interest:  Confinement, the existence and spectrum of bound states, and chiral symmetry breaking to name a few.  Studying these phenomena from the standpoint of quantum field theory is difficult, and impossible when confined to the realm of perturbation theory.

Recently, much progress has been made by constructing brane background models with a tunable parameter, such that in one limit they become pure (weakly coupled) $SU(N_c)$ QCD, and in the opposite they become a weakly coupled string theory \cite{Witten:1998zw}.  Such constructions allow one to study a model that is in the same universality class as pure YM theory  via gauge gravity duality \cite{Maldacena:1997re} (for a review see \cite{Aharony:1999ti}).   Karch and Katz \cite{Karch:2002sh} introduced a holographic dual of dynamical flavored fundamental quarks by adding  flavor  probe D-branes into the gravity background. By taking the number of flavor branes to be much smaller than $N_c$,  $N_f\ll N_c$, the backreaction on the background is avoided. This is the analog of the
  quenched approximation in  the gauge  theory. This was done originally for non-confining backgrounds. In  \cite{Sakai:2003wu} a similar scenario was proposed in a confining setup.\footnote{ For references of additional papers that discussed holographic
dynamical quarks see \cite{Erdmenger:2007cm} }
  The incorporation of chiral symmetry and its spontaneous breakdown  has been done in the seminal papers  of Sakai and Sugimoto (SS)\cite{Sakai:2004cn,Sakai:2005yt}. Their  model which has been extensively studied in the literature.

In the SS model, the background is taken to be a near extremal D4 with one of the spatial coordinates compactified, which we will call $x_4$, and its radius of compactification is $R_{x_4}$.  For energies $E\ll\frac{1}{R_{x_4}}$ one expects 4D dynamics contaminated with KK modes.  In the low temperature regime, the correct place to put the blackening factor is in front of the $dx_4^2$ component of the metric, giving the $u$,$x_4$ plane the geometry of a (semi-infinite) ``cigar''.  The probe flavor branes are taken to be ($N_f$) D8 branes, and are parallel to some combination of the radial direction, parameterized by $u$, and $x_4$.  In this model, the realization of chiral symmetry breaking $U(N_f)_R\times U(N_f)_L\rightarrow U(N_f)$ is geometric, and easy to visualize: the stack of D8 branes must connect smoothly at the bottom of the cigar, and so connect smoothly out to $\overline{\rm D8}$ branes at infinity, making a U-shaped profile in the $u$, $x_4$ plane. At large $u$, the UV, one sees a disconnected $D8$ $\overline{\rm D8}$ pair with $U(N_f)_R\times U(N_f)_L$ which is broken at small $u$, the IR, where one sees only the diagonal $U(N_f)$ symmetry remaining.  A variety of physical properties of meson
and baryon physics has been extracted from the model. These include the massive
meson spectrum, the massless Goldstone pions \cite{Sakai:2004cn}, certain decay rates \cite{Sakai:2005yt}, thermal behavior of hadrons \cite{Aharony:2006da}, \cite{Peeters:2006iu}, as well as describing certain condensation phenomena in the theory \cite{Bergman:2007pm,Dhar:2007bz,Dhar:2008um,Aharony:2007uu,Aharony:2008an,McNees:2008km,Argyres:2008sw}. It is clear, however, that the SS model is far from being the ultimate gravity dual of QCD. It suffers from several drawbacks like the fact that its UV completion is not known,
like discrepancies between its hadronic spectrum and that observed in nature. Thus, one is instructed to further construct
holographic laboratories. The present paper follows these lines.

Here, we will instead take a near extremal D5 background with two  of the world volume coordinates compactified ($x_4,x_5$), and take the flavor branes to be probe D7 branes.  We take probe D7 branes will be parallel to some combination of $u,x_4,x_5$; some aspects of this setup have been investigated in \cite{Edalati:2008xr}, and a close analog investigated in \cite{Antonyan:2006pg}.  We expect that the low energy dynamics to be once again 4D, and given the arguments of \cite{Witten:1998zw} (and applied to the D5 background in \cite{Antonyan:2006pg}), should also be in the same universality class as pure YM theory, excluding the flavor branes.  Including the flavor branes adds chiral fermions to the low energy weak coupling description, and so the holographic setup is dual to a theory in the same universality class as pure large $N_c$ QCD.

There are, however, some key differences between extremal D4 and D5 branes.  Most distinctly is the fact that the decoupling limit does not raise a large barrier, decoupling the bulk modes.  Rather, the radial effective potential becomes infinitely long, but of finite hight.  We find that this is true for the effective potential of the D7 brane fluctuations as well.  This gives that the effective potential has finitely many bound states, and so unfortunately only a finite number of 4D mesons.  We will, however, see the same type of geometric understanding of chiral symmetry breaking as described above.

To begin, we will now briefly review the near horizon limit of the near extremal D5 background we wish to study.  The metric is given by
\bea
ds^2&&=\frac{u}{R}\left(\eta_{\mu \nu} dx^{\mu}dx^{\nu}+dx_4^2+dx_5^2f(u;u_\Lambda)\right)+\frac{R}{u}\frac{du^2}{f(u;u_\Lambda)}+R u d\Omega_3^2
\eea
where $d\Omega_3^2$ is the metric of the unit 3 sphere, and
\be
f(u;u_\Lambda)=\left(1-\frac{u_\Lambda^2}{u^2}\right).
\ee
Furthermore, there is a non trivial dilaton and 3 form flux
\be
\exp{\phi}=g_s \frac{u}{R}, \quad F_3=\frac{2 R^2}{g_s} \Omega_3
\ee
where $\Omega_3$ is the volume form of the unit 3 sphere.  The parameter $R$ is related to string parameters in the usual way $R^2=g_s N_c \alpha'$.  We will be considering the case where both $x_4$ and $x_5$ are compact
\be
x_4=x_4+2\pi R_{x_4},\quad x_5=x_5+2\pi R_{x_5}.
\ee
The parameter $R_{x_5}$ is in fact not independent of the other scales already mentioned, given that we want the solution to be smooth around $u=u_\Lambda$.  One can see this most easily by changing to variables
\be
u^2=u_\Lambda^2+z^2.
\ee
In these variables, the metric reads
\bea
ds^2&&=\frac{\sqrt{u_\Lambda^2+z^2}}{R}\left(\eta_{\mu \nu} dx^{\mu}dx^{\nu}+dx_4^2\right)+R\sqrt{u_\Lambda^2+z^2} d\Omega_3^2+\frac{R\left(dz^2+z^2\frac{dx_5^2}{R^2}\right)}{\sqrt{u_\Lambda^2+z^2}}
\eea
and clearly, for the metric to be non singular around $z=0$, we need to take
\be
x_5=x_5+2\pi R, \quad \mbox{i.e.}\;\; R_{x_5}=R.
\ee
Thus we see that unlike the SS here $u_\Lambda$ is a free parameter not related to $R$.
In fact this is the situation classically. Based on similar situations \cite{Aharony:2004xn}, it is plausible that quantum   mechanically the system will not be stable for any $u_\Lambda$. However, for the case discussed in section 3 this instability does ont occur.
One should also note that the above background is confining, as $\sqrt{g_{tt} g_{xx}}$ is finite.

For this background, we read off the relation between field theory and string theory quantities
\bea
&& g_6^2=(2 \pi)^3 g_s \alpha', \quad g_4^2= \frac{2\pi g_s \alpha'}{R R_{x4}} \nn \\
&& T_{st}=\frac{1}{2\pi \alpha'} \sqrt{g_{tt} g_{xx}}|_{u=u_\Lambda}=\frac{u_{\Lambda}}{2\pi \alpha' R}.
\eea
The fact that   $u_\Lambda$ is a free parameter implies that so is the string tension.
Further, we should comment that this background is S-dual to the near extremal NS5 brane.  Such NS5 brane constructions have been argued to be the holographic dual of little string theories (LSTs) \cite{Aharony:1998ub} (for a review, see \cite{Kutasov:2001uf}), however in the case at hand we are in the opposite range of validity ($g_s \ll 1$).  Further, because S duality works non trivially on the D7 brane probes, our theory will not have a straightforward UV completion.  The use of holographic techniques for the NS backgrounds, \cite{Marolf:2007ys,Parnachev:2005hh,Elitzur:2000pq}, should, however, be easily be adapted to our scenario (and in some cases be identical).

We organize the remainder of the paper as follows.  In section 2 we discuss the general behavior for D7 embeddings in the confined phase, and briefly discuss the eigenvalue problem to describe the masses of mesons in this general case.  In section 3, we consider D7 embeddings into the extremal limit of the background, which is also sufficient to describe D7 embeddings that do not come near to the end of the radial coordinate at $u=u_\Lambda$.  We find that these embeddings obey a simple scaling behavior, which we describe generally enough to be applicable outside of these particular models.  We also consider fluctuations of the embedding functions and the world volume gauge field to describe the spectrum of mesons in this limit.  In section 4, we describe an embedding that falls all the way to $u=u_\Lambda$, and again consider fluctuations of the embedding functions and world volume gauge field.  We also consider baryons in this scenario.  Finally, in section 5 we consider the deconfined case, and consider the pattern of chiral symmetry breaking as a function of the boundary conditions.  In section 6 we conclude, and suggest some future lines of investigation.

\section{D7 embeddings}

\begin{figure}[ht!]
\centering
\includegraphics[width=.5\textwidth]{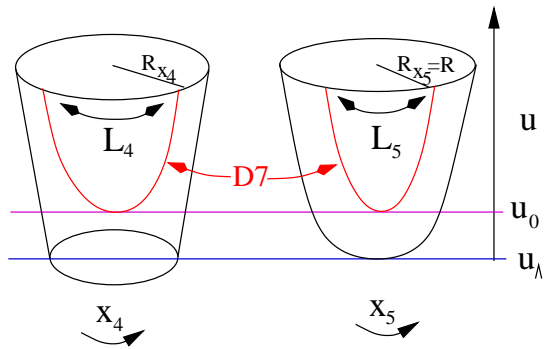}
\caption{The configuration of D7 branes in the near extremal D5 brane background, following the above considerations.}
\label{cigar}
\end{figure}

In this section we will consider embeddings of D7 branes into the above background.  Qualitatively, we will find embedding of the form shown in figure \ref{cigar}.
This background is similar to the paperclip or hairpin background (see for instance   \cite{Lukyanov:2003nj} and \cite{Kutasov:2004dj}).
The embeddings we consider are transverse to some combination of the $u,x_4,x_5$ directions, and so the Chern Simons (CS) term will be unimportant  in determining the solutions of the equations of motion, and we are left with the Dirac-Born-Infeld (DBI) action
\be
S_{D7}=-\kappa_7\int \d^8\xi \exp(-\phi)\sqrt{\left(-\det\left(g_{M N} \frac{\partial X^M}{\partial \xi^i}\frac{\partial X^N}{\partial \xi^j} \right)\right)} \label{dbi}
\ee
where we have used the notation $\xi^i$ to represent world volume coordinates of the D7 brane, and $\kappa_7$ is the D7 tension.
We use an Ansatz of the form
\bea
&& x^i(\xi)=\xi^i, \quad i=\mu, \quad i=\mbox{sphere coordinate} \\
 x_5(\xi)=x_5(\rho), \quad && x_4(\xi)=x_5(\rho), \quad u(\xi)= u(\rho). \nn
\eea
In the above notation, we have picked a ``static gauge'' for the $x^\mu$ and sphere directions.  The final world volume coordinate, $\rho$, parameterizes a path in $x_4, x_5, u$ space.  One can show directly that the full non linear equations of motion for the $x^\mu$ and sphere directions are satisfied for the above Ansatz.  Using the reparameterization invariance in $\rho$, we make one final choice
\be
u(\rho)=\rho. \label{gc1}
\ee
We now reduce the action to 1 dimension ($\rho$) and find the relevant equations of motion.  In the following we will abbreviate $\frac{\pa}{\pa \rho}=\pa$.  Using this notation, the action becomes
\be
S_{D7}=\frac{\kappa_7 V_3 V_4}{g_s} \int d\rho \rho^3 \sqrt{\left(\pa x_4\right)^2+\left(\pa x_5\right)^2f(\rho;u_\Lambda)+\frac{R^2}{\rho^2f(\rho; u_\Lambda)}}.
\ee
The equations of motion are quite simple, and follow from the Noether charges associated with translation invariance in $x_4$ and $x_5$.  These read
\bea
\frac{\rho^3 \pa x_4}{\sqrt{\left(\pa x_4\right)^2+\left(\pa x_5\right)^2f(\rho;u_\Lambda)+\frac{R^2}{\rho^2f(\rho; u_\Lambda)}}}=P_4 u_\Lambda^3 \label{defP4}\\
\frac{\rho^3 f(\rho;u_\Lambda) \pa x_5}{\sqrt{\left(\pa x_4\right)^2+\left(\pa x_5\right)^2f(\rho;u_\Lambda)+\frac{R^2}{\rho^2f(\rho; u_\Lambda)}}}=P_5 u_\Lambda^3 \label{defP5}.
\eea
This is a completely integrable system: one may simply solve for $\pa x_4$ and $\pa x_5$ and then integrate with respect to $\rho$.  Doing so, we find
\bea
x_4=\int d\rho P_4\sqrt{\frac{ \left(\frac{R}{u_\Lambda}\right)^2}{\left(\frac{\rho}{u_\Lambda}\right)^2f(\rho;u_{\Lambda})
\left(\left(\frac{\rho}{u_\Lambda}\right)^6-P_4^2-\frac{ P_5^2}{f(\rho;u_{\Lambda})}\right)}} \label{x4} \\
x_5=\int d\rho P_5 \sqrt{\frac{ \left(\frac{R}{u_\Lambda}\right)^2}{\left(\frac{\rho}{u_\Lambda}\right)^2f(\rho;u_{\Lambda})^3
\left(\left(\frac{\rho}{u_\Lambda}\right)^6-P_4^2-\frac{ P_5^2}{f(\rho;u_{\Lambda})}\right)}}. \label{x5}
\eea
The above expressions represent only one branch of the solution: in the end one must match an identical branch connecting at the minimum value of $u_0=\rho_0$.  One may check explicitly that the field equations of motion arising from action
(\ref{dbi}) are met for the solution given by (\ref{x4}), (\ref{x5}) and (\ref{gc1}).

There is a similarity to the solution for $x_5$ and the solution given in \cite{Aharony:2006da}, except that a different power of $\rho$ appears with the (also different) function $f$.  In fact, for the case $P_4=0$, we find that the minimum value of $u$ is in fact given by $u_0$, with $P_5=\frac{u_0^3}{u_\Lambda^3}\sqrt{f(u_0;u_\Lambda)}$ which is a very similar expression to that given in \cite{Aharony:2006da}.

The new piece of information is given by $P_4$, and this changes the situation in an interesting way.  First, we note that the ``turn around point'' for the embedding is given where $\pa x_4$ or $\pa x_5$ blow up (i.e. $u=\rho$ is not varying while both $x_4$ and $x_5$ vary allot).  Note that for both $x_4$ and $x_5$ this happens at the same place: either where $f(\rho;u_\Lambda)\rightarrow 0$ or where $\left(\left(\frac{\rho}{u_\Lambda}\right)^6-P_4^2-\frac{ P_5^2}{f(\rho;u_{\Lambda})}\right)\rightarrow 0$.  The question then becomes, which one goes to zero first?

To help address these questions, we first change to variables
\bea
\rh=\frac{\rho}{u_\Lambda}.
\eea
Under this transformation we are considering where
\bea
f(\rho; u_\Lambda)=f(\rh;1)=0 \nn \\
\left(\rh^6-P_4^2-\frac{ P_5^2}{f(\rh;1)}\right)=0.
\eea
We can now see which one goes to 0 first: it is clearly the second line (when $P_5\neq0$).  Imagine coming in from very large positive $\rh$ and decreasing this value.  The first line goes to zero only when $\rh=1$.   However, in such a limit, the second line already has a zero.  This is because of the $f$ in the denominator of the last term is going to zero (from the positive side), and so at some point this will compensate for the value of $\rh^6-P_4^2$, and have a zero at a value of $\rh > 1$.

The only special case is when $P_5=0$.  When this is the case, one of the functions goes to zero at ${\rm max}\left(P_4^{\frac 13},1\right)$.  Note that this has the peculiar feature that the endpoints of the D7 brane (a function of $P_4$) may not be at the same point, but none the less, the brane falls all the way to the bottom of the cigar $u=u_\Lambda$.  This ceases once $|P_4|>1$.  If $|P_4|>1$, then the embedding isn't antipodal any more in $x_5$.  In the $x_5,u$ plane, the embedding never reaches the bottom point, and so then never needs to be matched to a solution going through the origin (we will see this more clearly we plot the asymptotic separation distance as a function of $P_4$ and $P_5$).

\begin{figure}[ht]
\centering
\includegraphics[width=.45\textwidth]{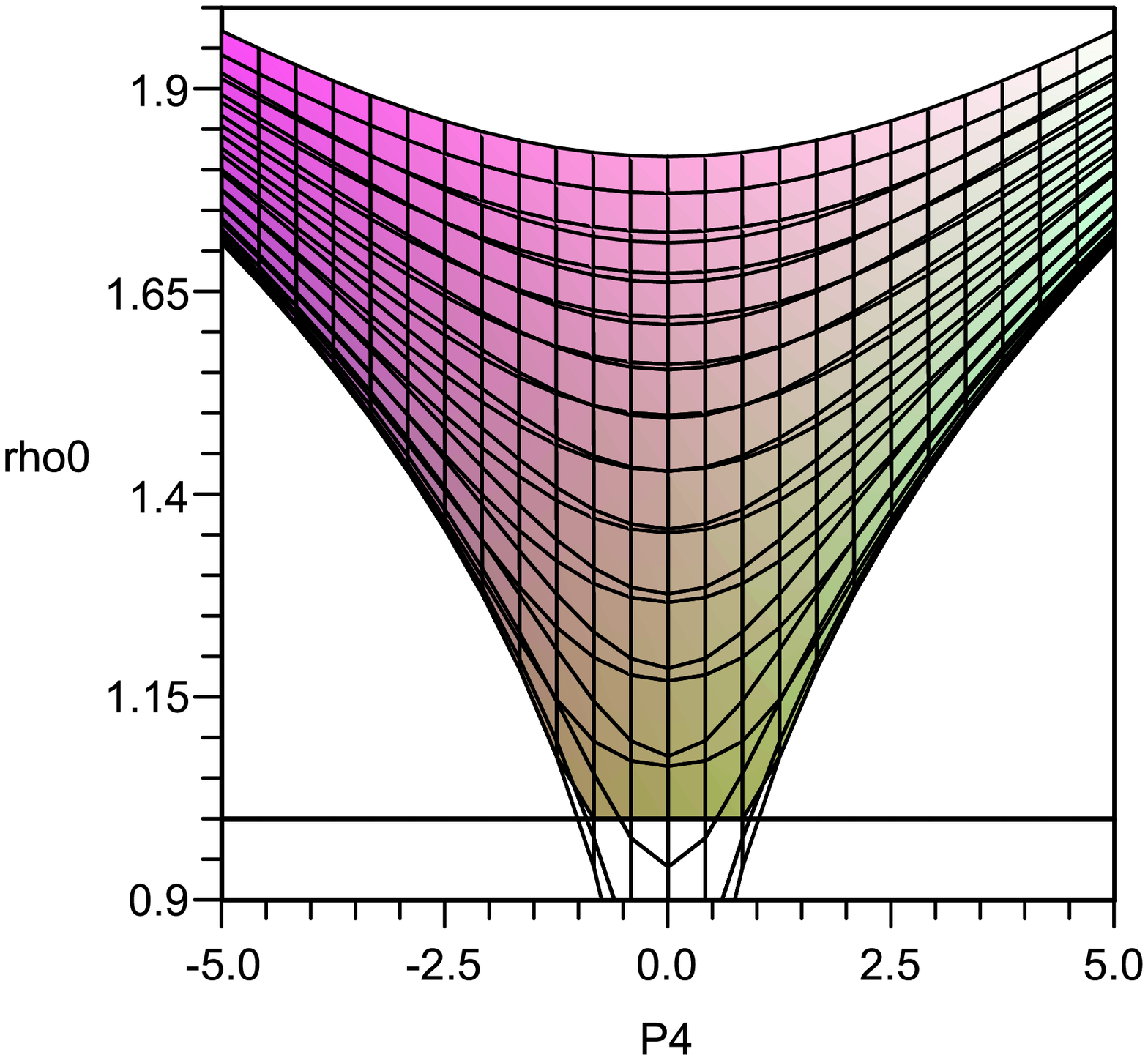}
\includegraphics[width=.45\textwidth]{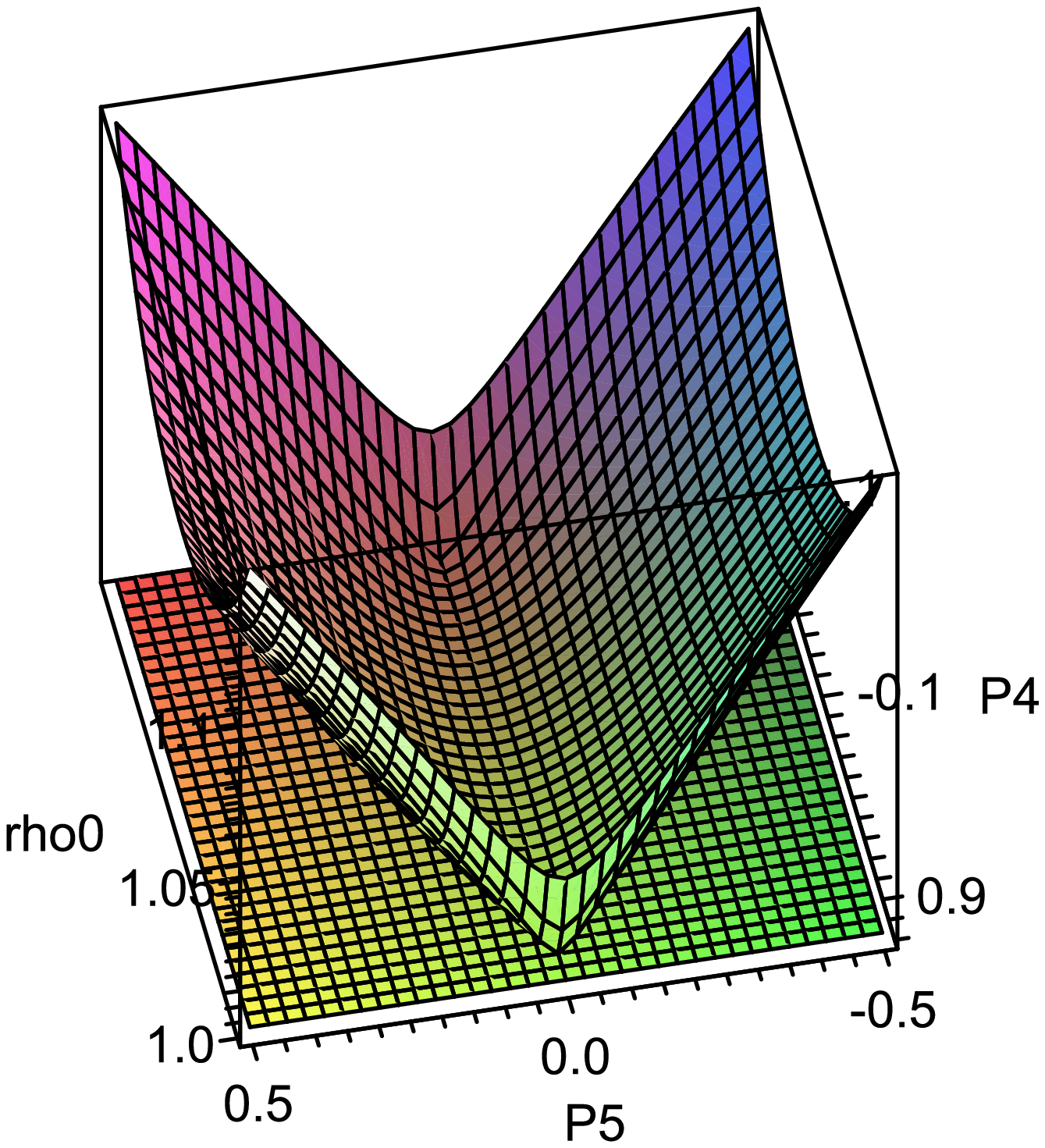}
\caption{The left hand graph is a side view of the plot of the root of the above rational function, plotted as a function of $P_4,P_5$.  The black wire frame plot is the asymptote $\rh_0=(P_4^2+P_5^2)^{\frac 16}=P_4^{\frac 13}$ when $P_5=0$.  The right hand graph is a plot of $\rh_0(P_4, P_5)$ zoomed in on the region $P_5=0, |P_4|<1$. One may see that this surface is smooth on the line interval $P_5=0, |P_4|<1$ except the endpoints $P_5=0, P_4=\pm 1$.}
\label{bigRootSide}
\end{figure}

One can check the above discussion of $u_0=\rho_0=\rh_0 u_\Lambda$ simply by plotting the results.  We define $\rh_0$ by the equation
\be
\left(\rh_0^6-P_4^2-\frac{ P_5^2}{f(\rh_0;1)}\right)=0.
\ee
\begin{figure}[ht]
\centering
\includegraphics[width=.45\textwidth]{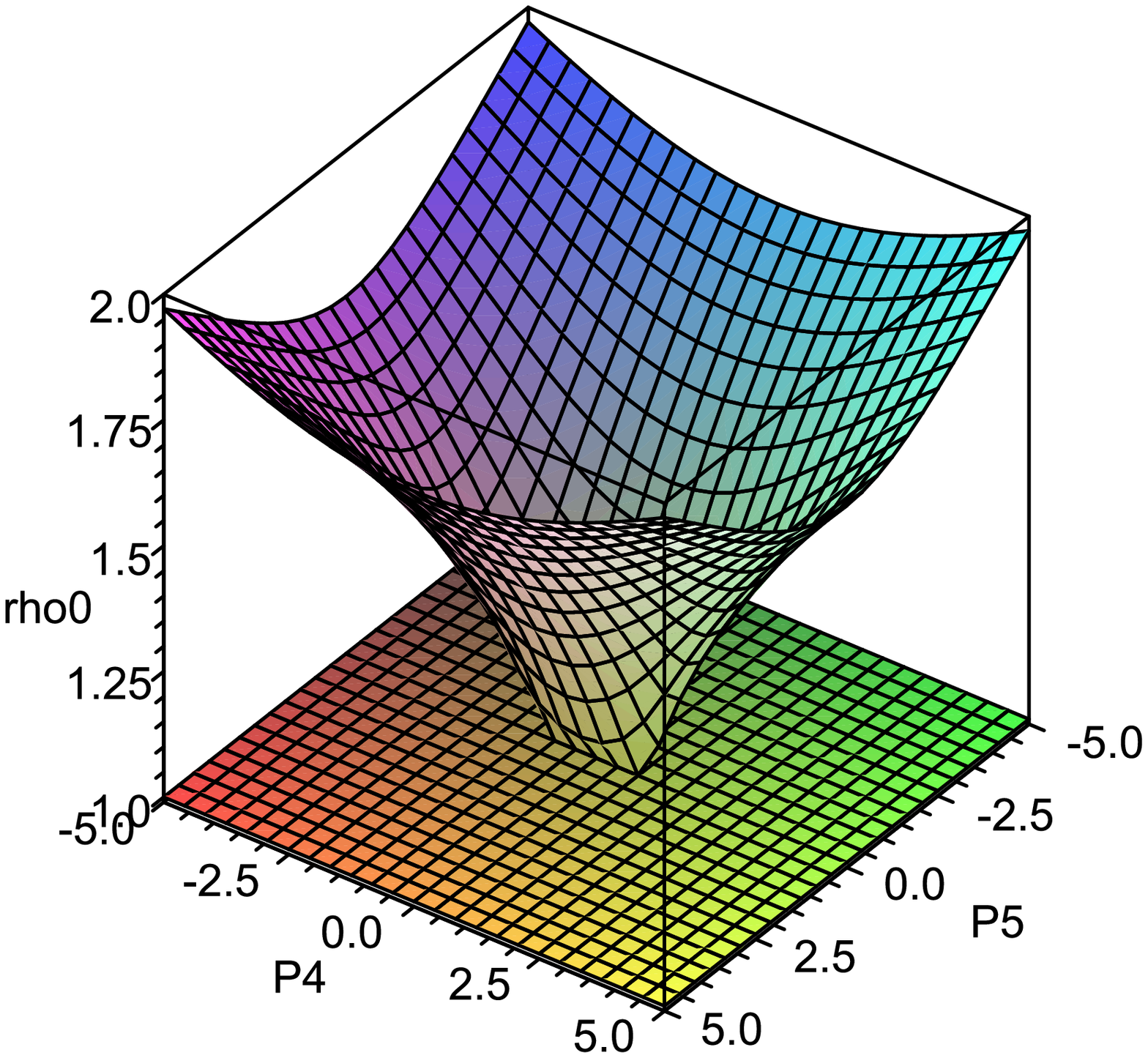}
\includegraphics[width=.45\textwidth]{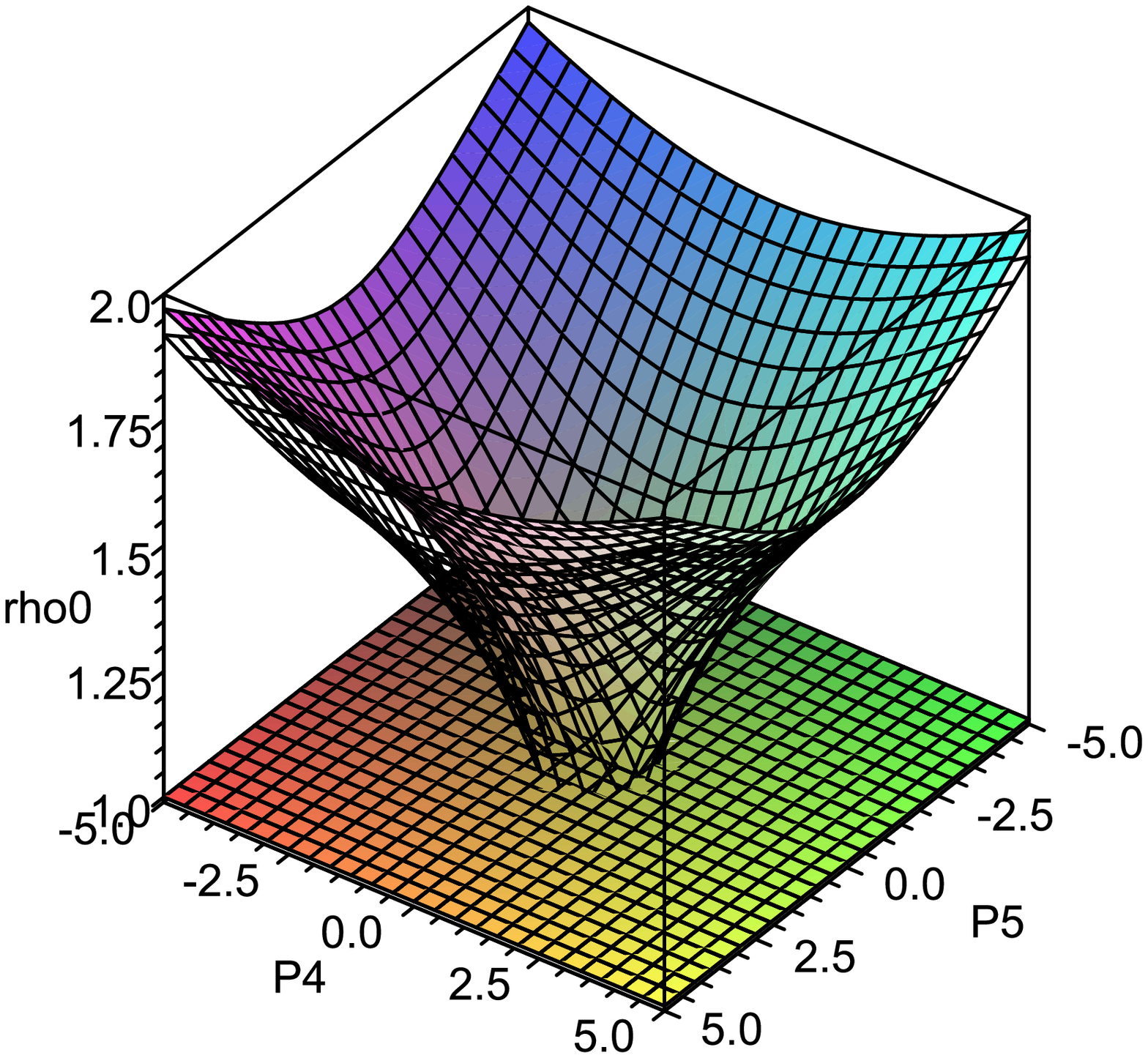}
\caption{A plot of the largest real root, $\rh_0(P_4,P_5)=u_0(P_4,P_5)$.  Note all values are larger than 1.  The wire frame in the right hand plot corresponds to the large $P_i$ asymptote $\rh_0(P_4,P_5)\rightarrow(P_4^2+P_5^2)^{\frac 16}$.}
\label{bigRootAsymp}
\end{figure}
Clearly, $\rh_0$ is a function $\rh_0(P_4,P_5)$ and so we may plot this (we plot the largest real root), and compare it with the value $1$ in figure \ref{bigRootAsymp}.
Further, we can see the special case of $P_5=0$ arising, and the flattening of the function that happens along the line interval $|P_4|<1$: indeed ${\rm max}\left(P_4^{\frac 13},1\right)$ is the limiting value of the solution along the line $P_5=0$.  Further, for large values of $P_4$ and $P_5$, the value of the zero $\rh_0$ grows as well, and so the $f$ in the last part of the expression can be approximated by 1.  This gives that the approximate value of the zero is given by $\rh_0=(P_4^2+P_5^2)^{\frac 16}$.  We plot this limiting value and show that it does indeed asymptote (see figure \ref{bigRootAsymp}).  Finally, one can show that this surface exactly corresponds to the curve ${\rm max}(P_4^{\frac 13},1)$ along the line $P_5=0$ by plotting a side view (see figure \ref{bigRootSide}).

One may be curious how $\rh_0\rightarrow 1$ as $P_5\rightarrow 0$ on the line interval $|P_4|<1$.  Clearly, it must relax as $P_5^2$, and so must be smooth around $P_5=0$.  We plot a zoomed in version of the plot of $\rh_0$
and find that this is indeed the case.  The only singular points on the surface (with conical singularities) are at $(P_5,P_4)=(0,\pm 1)$.  See figure \ref{bigRootSide}.

\begin{figure}[ht!]
\centering
\includegraphics[width=.4\textwidth]{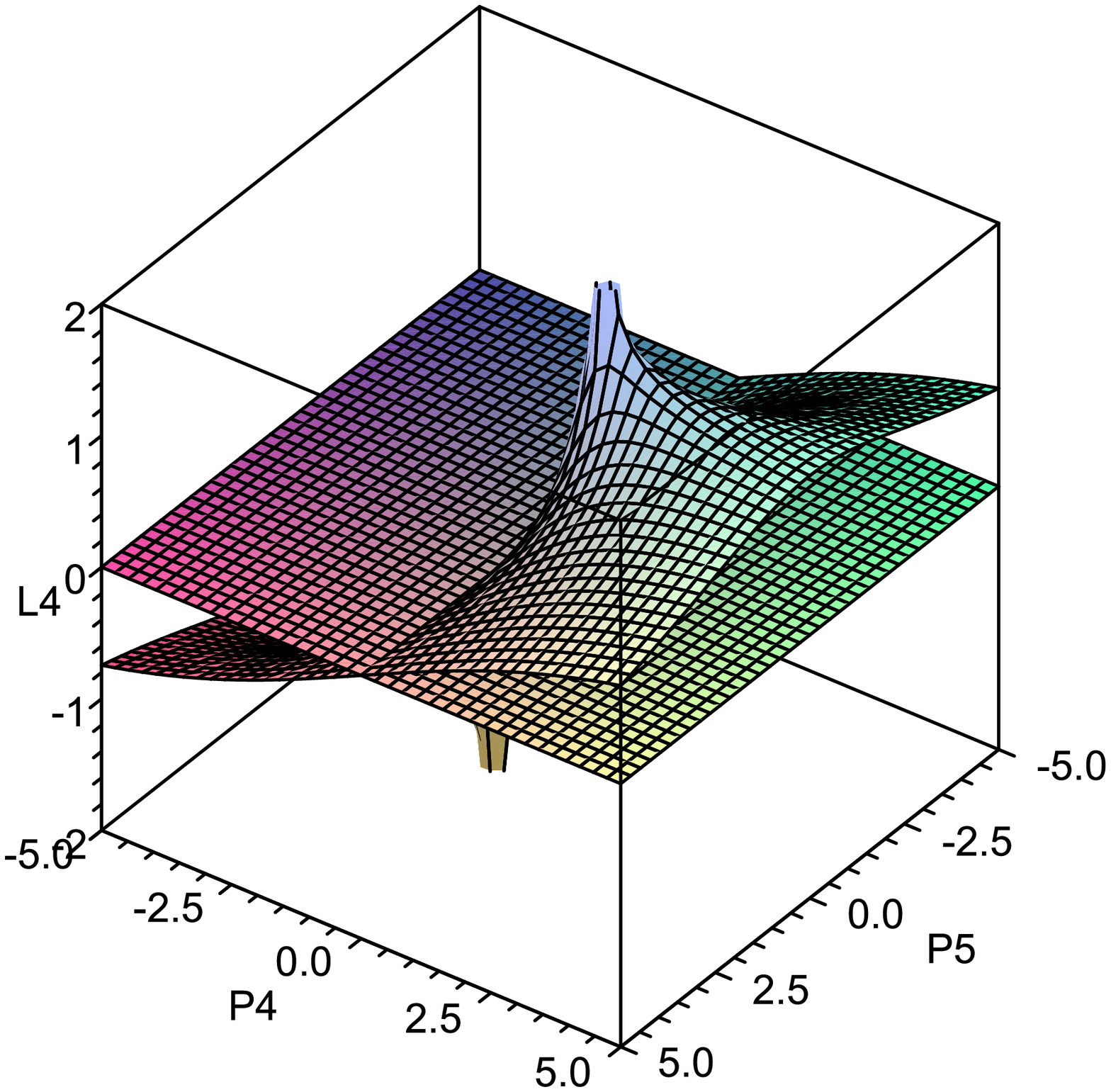}
\includegraphics[width=.4\textwidth]{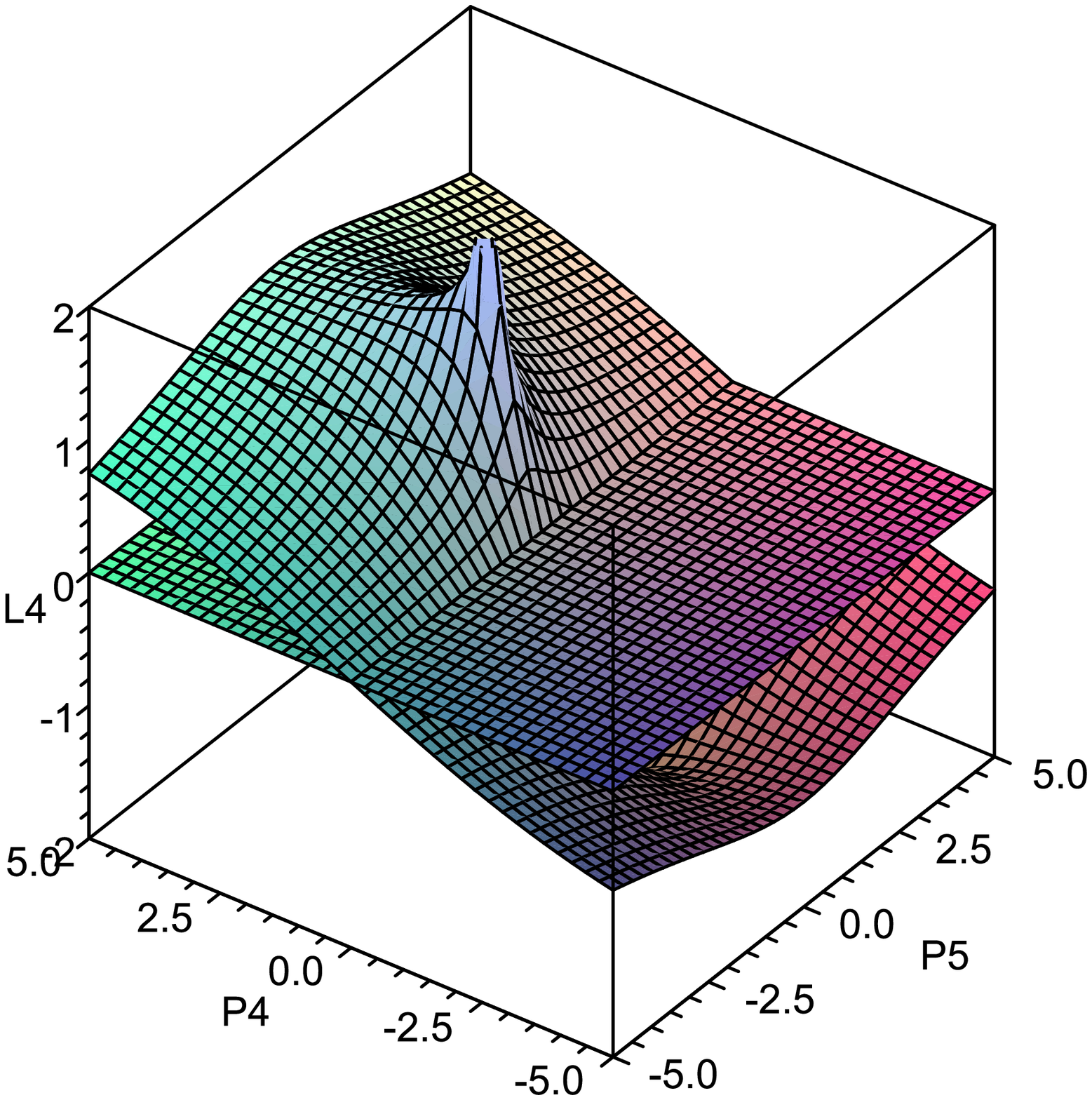}
\includegraphics[width=.4\textwidth]{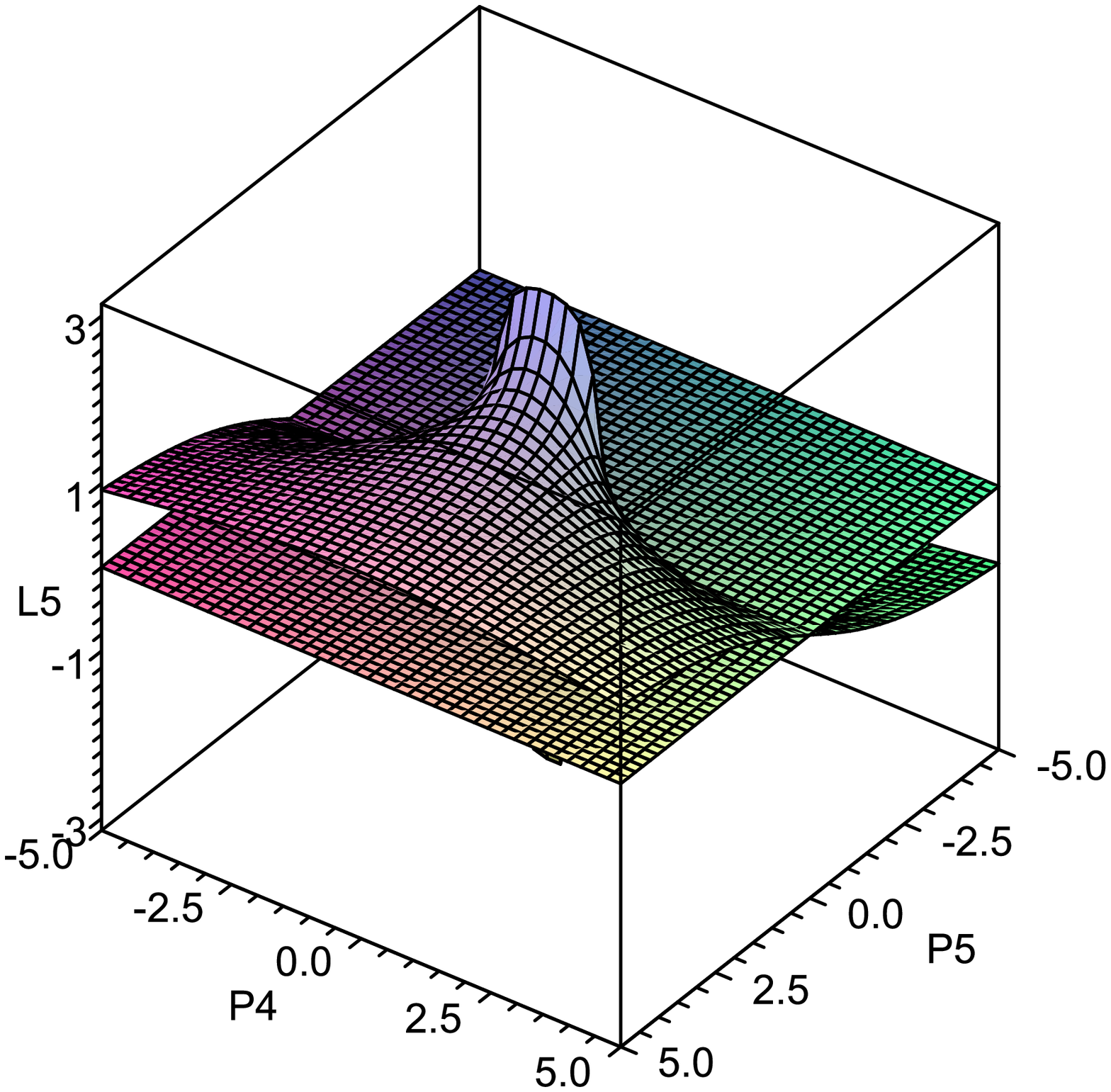}
\includegraphics[width=.4\textwidth]{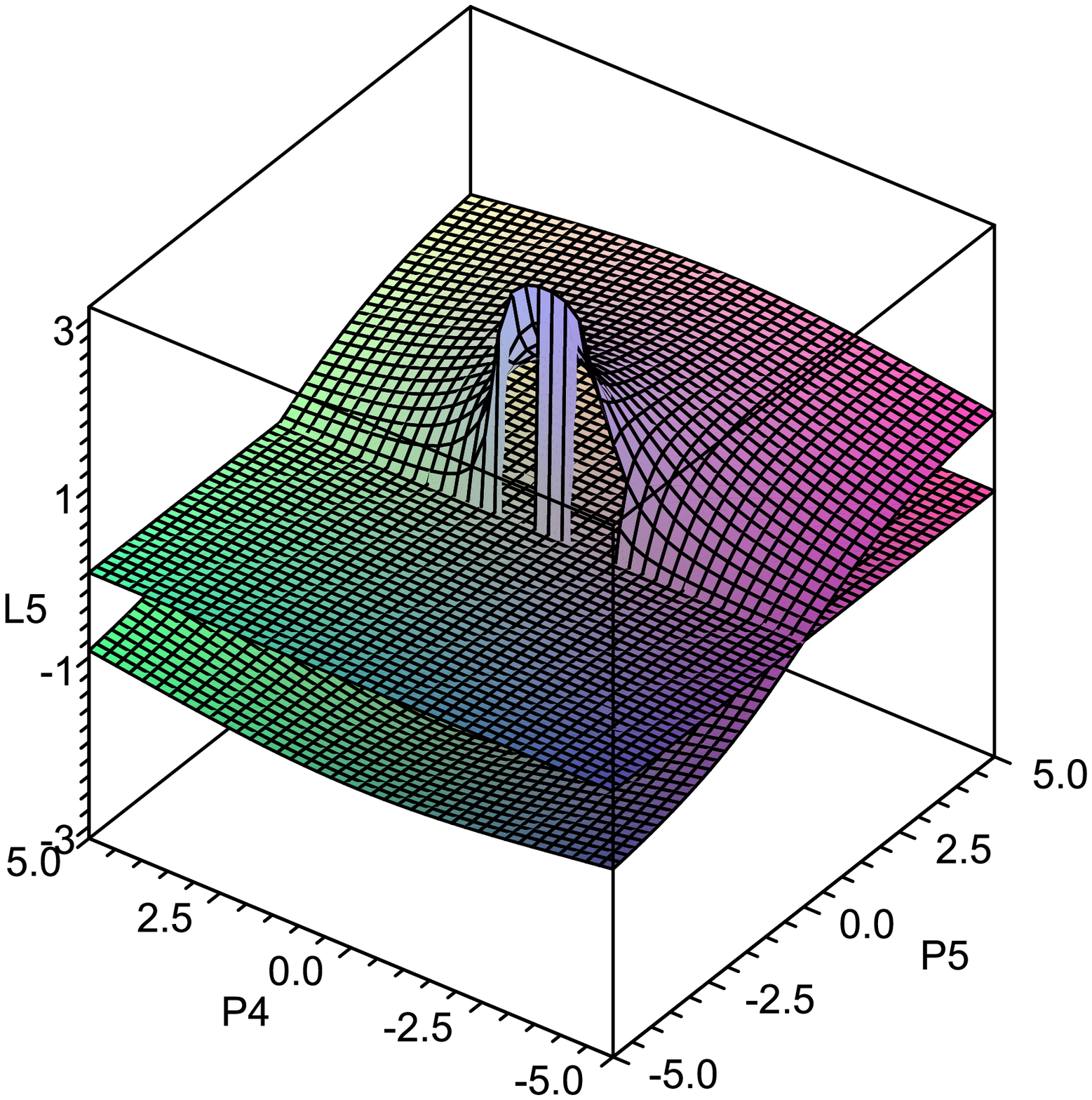}
\caption{Plots of $L_4(P_4,P_5)/R$ and $L_5(P_4,P_5)/R$ (labeled simply L4 and L5 in the graphs).  The view on the left gives the same perspective as earlier graphs, while the view on the right is slightly easier to see what is happening near the surface $L_4(P_4,P_5)/R=0$ and $L_5(P_4,P_5)/R=0$.}
\label{L4L5}
\end{figure}

Now that we have found (qualitatively) the minimum value to which the brane falls, we also wish to relate the parameters $P_4$ and $P_5$ to brane separation at $u=\infty$ (in the coordinates $x_4$ and $x_5$).  We will call the asymptotic separation in $\Delta x_4= L_4$ and $\Delta x_5= L_5$.  In the end it will be these separations that we wish to consider the ``boundary conditions'' imposed on the branes, so that the fluctuations around these embeddings will satisfy Dirichlet boundary conditions.  Given the above expressions, we see that
\bea
L_4=2R\int_{\rh_0}^{\infty} d\rh P_4\sqrt{\frac{1}{\rh^2 f(\rh;1)
\left(\rh^6-P_4^2-\frac{ P_5^2}{f(\rh;1)}\right)}} \\
L_5=2R\int_{\rh_0}^{\infty} d\rh P_5 \sqrt{\frac{1}{\rh^2f(\rh;1)^3
\left(\rh^6-P_4^2-\frac{ P_5^2}{f(\rh;1)}\right)}}.
\eea
where the factor of 2 comes from ``gluing'' the second branch of the solution on.  Again, finding analytic solutions to the above is problematic, but we can graph the results as a function of $P_4$ and $P_5$, see figure \ref{L4L5}.

\begin{figure}[ht]
\centering
\includegraphics[width=.4\textwidth]{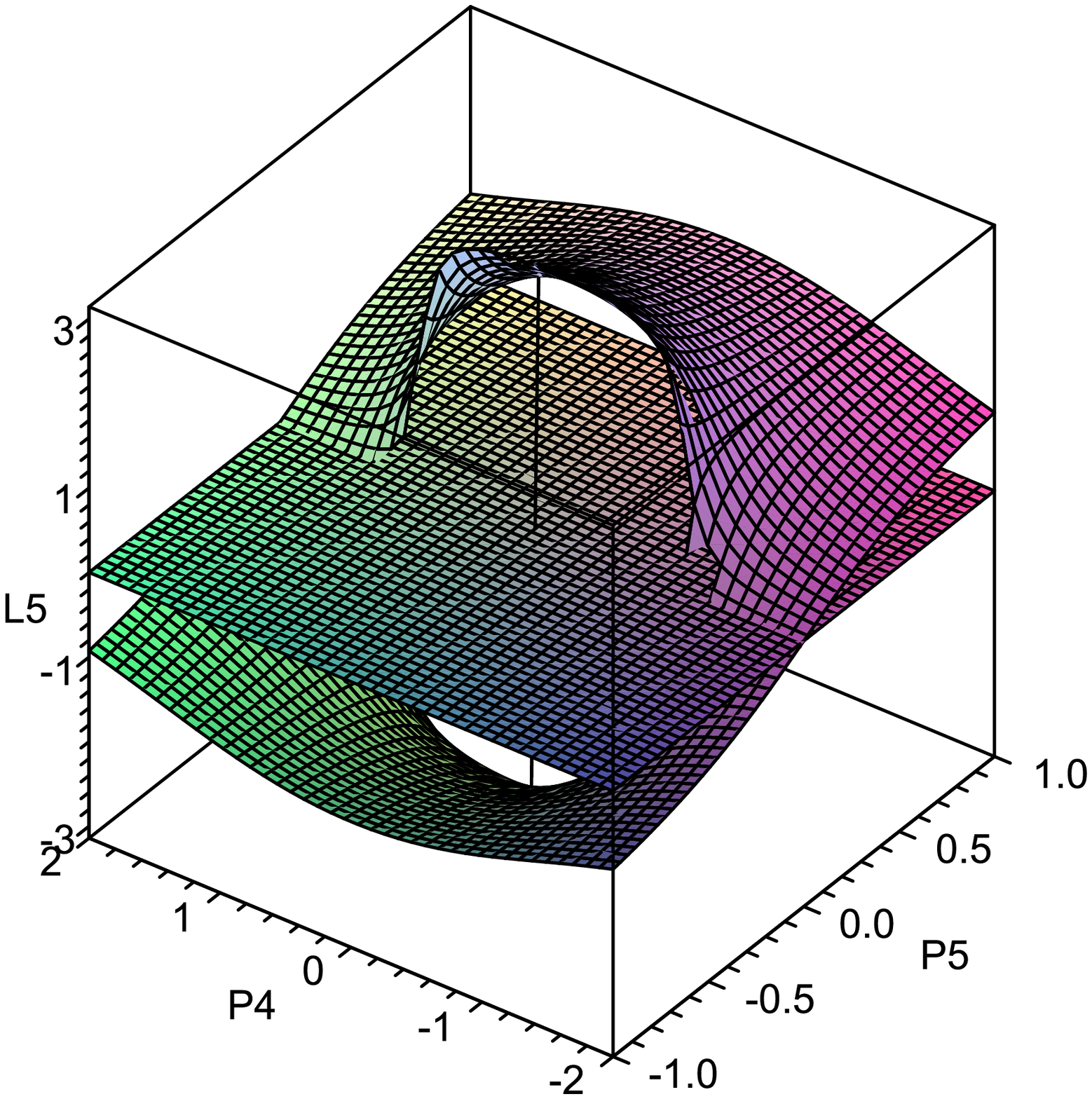}
\includegraphics[width=.4\textwidth]{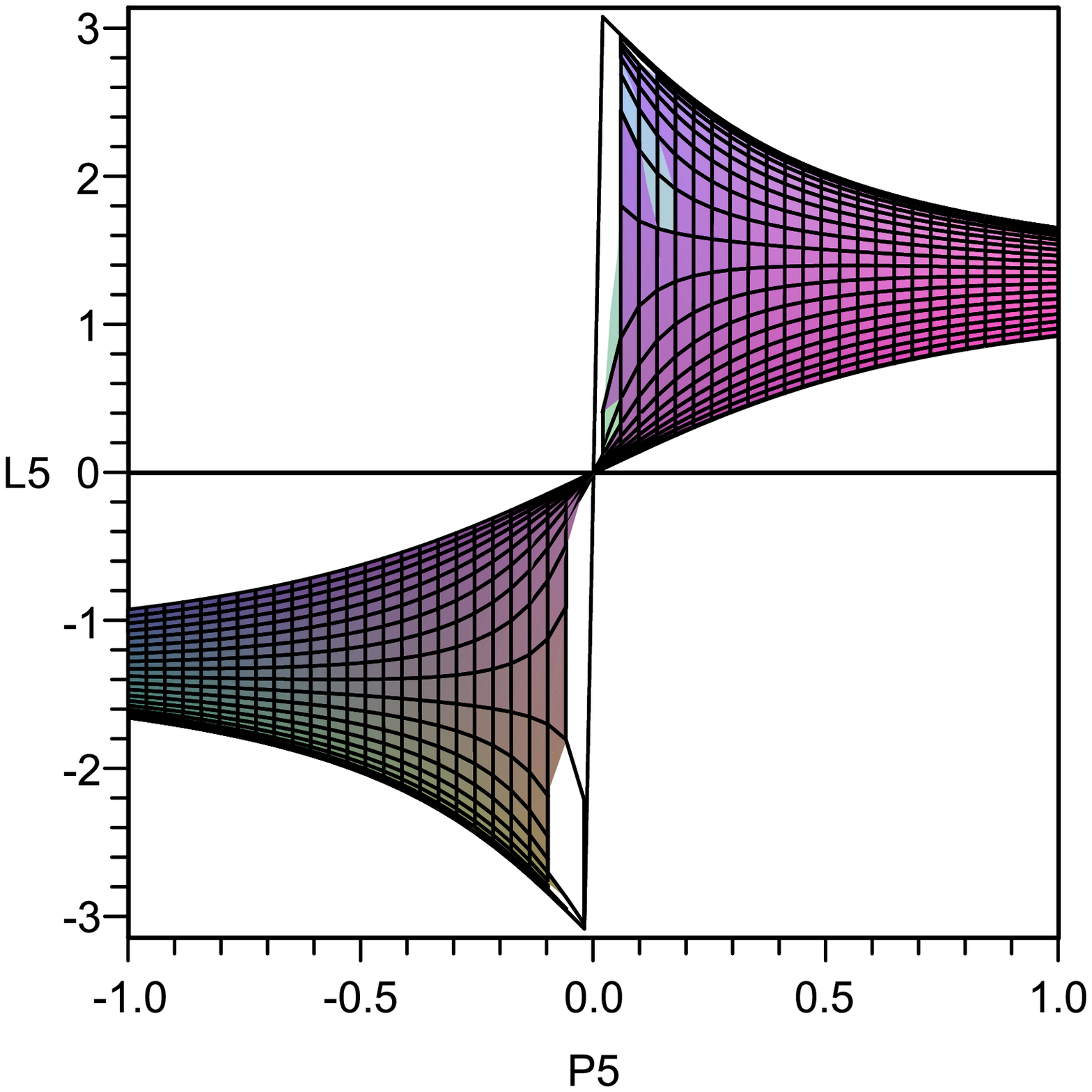}
\caption{Plots of $L_5(P_4,P_5)/R$ zoomed in around the interval.  Note that in the side view the maximum and minimum appears to approach $\pm \pi$. }
\label{L5Zoom}
\end{figure}

We first consider the graph of $L_5/R$.  Before making this graph, one may think that we need to be careful about finding the actual separation distance $L_5$: one could have a solution that connects through smoothly the point $u=u_\Lambda$, and so the second branch actually lies displaced by $\pi R_{x_5}=\pi R$.  However, the integral for $L_5/R$ appears to have this built in.  The only cases where the turn around point of the branes lies at $u_0=u_\Lambda$ is on the interval $-1<P_4<1$ and $P_5=0$, and so we focus on this region.  While we do not offer an analytic proof, we may zoom in on this part of the graph, and be convinced that $L_5/R\rightarrow \pm \pi$ (see figure \ref{L5Zoom}).  One may also be convinced simply by noting that if $P_4=P_5=0$, then the second branch must be glued on in antipodally after reaching $u=u_\Lambda$ to avoid a discontinuity in the tangent vector in $u,x_5,x_5$ space.  If this is the case, we see that the integrand in the expression for $L_5$ must collapse to a delta function with coefficient $\pm \pi$ in the limit that $P_5\rightarrow \pm 0$.  For generic values of $\rh$ the limit $P_5\rightarrow 0$ sets the integrand to zero, however at $\rh=\rh_0$ it blows up to infinity with the integral underneath remaining constant.  This is sufficient to show that the integrand goes to a delta function in $\rh$ for the limit $P_5\rightarrow 0$.  This also explains why one misses this if one passes the limit $P_5=0$ under the integration: the limit implied by the integral and the limit $P_5\rightarrow 0$ do not commute, and so one must put in the additional $\pi$ of separation by hand. Further, because the values $\pm \pi$ are identified, the surface is actually continuous.

Next we consider the graph for $L_4$.  In figure \ref{L4L5} there is a divergence in $L_4$ around $P_5=0$ and $P_4\rightarrow \pm1$.  This is easy to understand.  If one looks at $P_5=0$ and $P_4=1$, the integrand goes as $\frac{1}{\rh -1}$ around $\rh=\rh_0=1$, and so the integral is log divergent.
Because $L_4$ blows up around $P_5=0, P_4=\pm1$ we may wish to exclude multiple wrappings of the D7 around $x_4$.  We further restrict the total $x_4$ winding to be $\pi/R_{x_4}$ (half the circumference), knowing that there is another solution where the same boundary conditions are satisfied, and less distance has been traversed by the brane.  Hence, one may wish to exclude a small region around $P_5=0$ and $P_4= \pm1$ whose boundary is given by $|L_4|/R =\pi R_{x_4}/R$.

The above considerations are what lead to the diagram in figure \ref{cigar}.

\subsection{Mesons in the model.}

The mesons of this model are described by strings ending on the D7 branes\cite{Kruczenski:2004me}.  The high spin mesons are given by semiclassical strings which end on the D7 at the minimal value of the brane embedding $u_0$, fall towards the minimum value of $u_\Lambda$ and lay flat at this value and finally climb back up to meet the D7 again at $u_0$.  The low spin (vector) mesons are given by fluctuations of the D7 $U(N)$ gauge field strength, and the scalar mesons are given by the fluctuations of the embedding functions.

To compute any of these fluctuations, we will need the pullback metric
\bea
ds_p^2&&=\frac{\rho}{R}\left(\eta_{\mu \nu} dx^{\mu}dx^{\nu}+\left(\pa x_4^2+\pa x_5^2f(\rho;u_\Lambda)+\frac{R^2}{\rho^2}\frac{1}{f(\rho;u_\Lambda)}\right)d\rho^2+R^2 d\Omega_3^2\right) \nn \\
&&=\frac{u_\Lambda}{R}\rh\left(\eta_{\mu \nu} dx^{\mu}dx^{\nu}+\gamma(\rh) \frac{R^2}{\rh^2} d\rh^2+R^2 d\Omega_3^2\right).
\eea
where we have defined the useful function
\be
\gamma(\rh)\equiv \frac{\rh^6}{f(\rh;1)\left(\rh^6-P_4^2-\frac{P_5^2}{f(\rh;1)}\right)}.
\ee
For later use, we will also need
\be
\exp(-\phi)\sqrt{-\det\left(g_{(p) ab}\right)}=\frac{u_\Lambda^3 R}{g_s}\rh^2 \sqrt{\gamma(\rh)}
\ee
and we find it convenient to use the notation
\be
\frac{\pa}{\pa \rho}= \pa, \quad \frac{\pa}{\pa \rh}=\pah=u_\Lambda \pa,\quad \frac{\pa}{\pa x^{\mu}}=\pa_\mu.
\ee
With such an assignment, the solutions above read as
\bea
\pah x_4=R P_4 \frac{1}{\sqrt{\rh^2f(\rh;1)\left(\rh^6-P_4^2-\frac{ P_5^2}{f(\rh;1)}\right)}}=\frac{RP_4}{\rh^4}\sqrt{\gamma(\rh)}  \label{dhx4}\\
\pah x_5=R P_5 \frac{1}{\sqrt{\rh^2f(\rh;1)^3
\left(\rh^6-P_4^2-\frac{ P_5^2}{f(\rh;1)}\right)}}=\frac{RP_5}{\rh^4 f(\rh;1)}\sqrt{\gamma(\rh)}\label{dhx5}.
\eea

\subsubsection{Vector mesons}

First, we compute the effective action for the fluctuations associated with the gauge field $F_{a b}$ which we take to be only a function of $\rh , x^{\mu}$ and further that components of $F$ with indices along the sphere directions are zero.
With this, we take the DBI and expand it to second order in F \footnote{We need not worry about the CS term because the pullback of $A_6$ (generated by the $D5$ background) onto the D7 worldvolume is zero.  Note also that we are taking $F$ to be a U(1) gauge field.  One may generalize this discussion to a U(N) field by promoting $F$ to a matrix, and taking a trace.}
\bea
S_{D7}&& =-\kappa_7\int \d^8\xi \exp(-\phi)\sqrt{-\det\left(g_{(p) ab}+F_{ab}\right)} \nn \\
&&=-\kappa_7\int \d^8\xi \exp(-\phi)\sqrt{-\det\left(g_{(p) ab}\right)} \nn \\
&& \quad \times\sqrt{\left(1+g_p^{a_1 b_1}F_{a_1 b_1}+\frac12 F_{a_1 b_1} F_{a_2 b_2} \left(g_p^{a_1 b_1}g^{a_2 b_2}-g_p^{a_1 b_2}g_p^{a_2 b_1}\right)+\cdots \right)}
\eea
and now because $F$ is antisymmetric, while $g$ is symmetric, we find that to leading order
\bea
&&=-\kappa_7\int d\Omega_3 \int dx^\mu d\rh \frac{u_\Lambda^3 R}{g_s}\rh^2 \sqrt{\gamma(\rh)} \left(1+\frac14 F_{a_1 b_1} F_{a_2 b_2} g_p^{a_1 a_2}g_p^{b_1 b_2}+\cdots \right).
\eea
Hence, the quadratic action we wish to consider is
\bea
S_{F^2}&&=-\frac{\kappa_7 \Omega_3}{4 g_s} \int dx^4 d\rh u_\Lambda^3 R\rh^2 \sqrt{\gamma(\rh)} \left(\frac{R^2}{\rh^2 u_\Lambda^2}F_{\mu_1 \mu_2} F^{\mu_1 \mu_2}
+\frac{2}{u_\Lambda^2 \gamma(\rh)}F_{\mu_1 \rh} F^{\mu_1}{}_{\rh} \right) \nn \\
&&=-\frac{\kappa_7 \Omega_3}{4 g_s} \int dx^4 d\rh u_\Lambda R^3\left(\sqrt{\gamma(\rh)}F_{\mu_1 \mu_2} F^{\mu_1 \mu_2}
+\frac{2\rh^2}{R^2 \sqrt{\gamma(\rh)}}F_{\mu_1 \rh} F^{\mu_1}{}_{\rh} \right)
\eea
and in the above we have used $\eta^{\mu \nu}$ to raise and lower indices.

The equations of motion following from the above action are
\bea
\sqrt{\gamma}\pa_\nu F^{\nu \mu}+\pah \sqrt{\gamma} F^{\rh \mu}=0 &\leftrightarrow&  \sqrt{\gamma}\pa^\nu F_{\nu \mu}+\pah\left(\frac{\rh^2}{R^2\sqrt{\gamma}}F_{\rh \mu}\right)=0 \label{Amu}\\
\pa_\mu F^{\mu \rh}=0 &\leftrightarrow& \pa^\mu F_{\mu \rh}=0.\label{Arh}
\eea
We expand the gauge field in the following way
\bea
A_\mu = \sum_n B^{(n)}_{\mu}(x^\nu) \psi_n(\rh), \quad A_{\rh} =\sum_n \pi_{(n)}(x^\mu) \phi_n(\rh).
\eea
We  further gauge fix $\pa^\mu B^{(n)}_\mu=0$.
Under these conditions, and for the above gauge choice, we  see that the equation (\ref{Arh}) becomes
\be
\sum_n m_n^2 \pi_n \phi_n(\rh)-\sum_{n}\pa_\mu B^\mu \pah \psi_n=\sum_n m_n^2 \phi_n(\rh)=0
\ee
where $m_n$ is defined by $\pa_\mu \pa^\mu B_\nu = m_n^2 B_\nu$.  We see that $\phi_n=0$ by this equation of motion except for the case where $m_n=0$.  We choose to index such a mode $m_0=0$.  $\pi_0$ is the massless pion.

We now note that the remaining equation of motion (\ref{Amu}) reads
\be
\sum_n\left[ \gamma^{\frac12}\pa_\mu \pa^\mu B_{(n)}^\nu \psi_n + \pah \left(\frac{\rh^2}{R^2 \gamma^{\frac12}} \pah \psi_n\right)B_{(n)}^\nu \right]- \pah \left(\frac{\rh^2}{R^2 \gamma^{\frac12}} \pa^\nu \pi_0 \phi_0 \right)=0.
\ee
One can show that there are no massless normalizable modes of the vectors $B_\nu$, and so we find that
\be
\pah \left(\frac{\rh^2}{R^2 \gamma^{\frac12}} \pa^\nu \pi_0 \phi_0 \right)=0 \rightarrow \phi_0\propto \frac{ \gamma^{\frac12}}{\rh^2}.
\ee
Thus, we find that the eigenvalue problem for the $\psi_n$ reduces to
\be
\frac{1}{\gamma^{\frac12}}\pah\left(\frac{\rh^2}{\gamma^{\frac12}}\pah \psi_n\right)=-M_{A,n}^2\psi_n
\ee
where the 4D mass is given by $m_{A,n}=\frac{M_{A,n}}{R}$.

Unfortunately, the eigenvalue problem for the fluctuations of the embedding coordinates are much more involved because we do not have a good set of orthogonal coordinates to the brane.  However, we will study these in two special cases below.

\section{Large $u_0\gg u_\Lambda$ limit.}

In this section, we consider the limit where $u_0\gg u_\Lambda$.  This limit is the same as the extremal limit of the background because $f(u;u_\Lambda)$ can be approximated by $f(u;u_\Lambda)=1$.
 As mentioned in the introduction this case makes sense also quantum mechanically. In fact it is similar to the cases studied in \cite{Antonyan:2006vw}.
 In such a limit the periodic identification of $x_4$ and $x_5$ now both become arbitrary, and further there is an additional $SO(2)$ symmetry rotating the $x_4,x_5$ plane, which the solutions will transform under in a simple way.  To consider the limit $u_0\gg u_\Lambda$, we cannot take a strict $u_\Lambda \rightarrow 0$ limit of equations (\ref{defP4}) and (\ref{defP5}) because this would scale the right hand side of these equations to zero (when they should be constants).  Therefore, we absorb the $u_\Lambda^3$ into $P_i$, via $p_i \equiv u_\Lambda^3 P_i$ (the parameters $p_i$ have units).  One may think of this as a scaling where $u_\Lambda\rightarrow 0$ while $P_i\rightarrow \infty$ with $p_i=u_\Lambda^3 P_i$ remaining fixed.  This is the connected to the fact that for large $P_i$ the solution is in a regime where $u_0\gg u_\Lambda$.

We consider equations (\ref{defP4}) and (\ref{defP5}) with the above substitutions and set $u_\Lambda= 0$ and find
\bea
\frac{\rho^3 \pa x_4}{\sqrt{\left(\pa x_4\right)^2+\left(\pa x_5\right)^2+\frac{R^2}{\rho^2}}}=p_4  \label{defsmallp4}\\
\frac{\rho^3 \pa x_5}{\sqrt{\left(\pa x_4\right)^2+\left(\pa x_5\right)^2+\frac{R^2}{\rho^2}}}=p_5  \label{defsmallp5}.
\eea
As before, we solve for $\pa x_4$ and $\pa x_5$, and find
\bea
x_4=\int d\rho R p_4\sqrt{\frac{1}{\rho^2
\left(\rho^6-p_4^2-p_5^2\right)}} = \frac{1}{3}\frac{R p_4 \arcsin\left(\frac{\sqrt{p_4^2+p_5^2}}{\rho^3}\right)}{\sqrt{p_4^2+p_5^2}}+x_{4,0}\\
x_5=\int d\rho R p_5\sqrt{\frac{1}{\rho^2
\left(\rho^6-p_4^2-p_5^2\right)}} =\frac{1}{3}\frac{R p_5 \arcsin\left(\frac{\sqrt{p_4^2+p_5^2}}{\rho^3}\right)}{\sqrt{p_4^2+p_5^2}}+x_{5,0}.
\eea
where the constants $x_{4,0}$ and $x_{5,0}$ can be fixed to 0 by translation invariance in $x_4$ and $x_5$ (again, the above only represents 1 branch of the solution and must be glued on to another branch).  This gives simple expressions for $L_4$ and $L_5$
\bea
L_4 &=& \frac{R p_4 \pi }{3 \sqrt{p_4^2+p_5^2}} \nn \\
L_5 &=& \frac{R p_5 \pi }{3 \sqrt{p_4^2+p_5^2}}.
\eea
Further, we should note that this gives that
\be
L_4^2+L_5^2=\frac{R^2 \pi^2}{9}.
\ee
In the $x_4, x_5$ plane, the ends of the D7 brane have a fixed asymptotic separation (not including the warp factor); similar behavior was found in \cite{Antonyan:2006pg} for D5 brane embeddings.  This gives that $L_4$ and $L_5$ do not parameterize the 2 dimensional class of solutions that we have.  We can, however, use $\rho_0=u_0=\left(p_4^2+p_5^2\right)^{\frac16}$ and the unit vector defining the separation direction $L_4/L_5=p_4/p_5$ as the two physical quantities.  Note that one can take fixed $L_4$ and $L_5$ while varying $u_0$ simply by rescaling $p_4$ and $p_5$ by the same factor.

Before turning to a discussion of these embeddings, we note some general points about how to find the scaling of $u_0$ with various $L_i$ in the problem.  We note that in the extremal limit, the action becomes
\be
S_{D7}=\frac{\kappa_7 V_3 V_4}{g_s} \int d\rho u^3 \sqrt{\left(\pa x_4\right)^2+\left(\pa x_5\right)^2+\frac{R^2 \pa u^2}{u^2}}
\ee
where in the above we have not fixed $u=\rho$ as a gauge choice.
We note that there is a Lie point symmetry (for more on Lie point symmetries, see \cite{Stephani}) of the equations of motion following from the above action.  Namely, if we take
\be
u \rightarrow C\times u
\ee
the action scales as
\be
S_{D7}\rightarrow C^3 S_{D7}.
\ee
This is clearly a symmetry of the solution space of the equations of motion.  The above symmetry is therefore a local map from solutions to solutions, the definition of a Lie point symmetry.

One could further scale the embedding coordinates $x^\mu(\xi)\rightarrow C^{n_\mu}x^\mu(\xi)$ (no summation) if one wishes (this rescales the effective $V_4$) and possibly promote the above symmetry to a Noether symmetry, however, we will not do this here \footnote{This is because the particular form of this symmetry is dependent on the fact that at least 4 of the fields $x^\mu,x^4,x^5$ are along the world volume of the D7 and aligning them to be the first 4 $x^\mu$ via a Lorentz transformation, and then further gauge fixing these to be $\xi^\mu$.  Thus, if we truly wanted to extend the above symmetry to a Noether symmetry, we would need to include some combination of coordinate transformations and Lorentz symmetry too, which we do not wish to consider here.}.
Note that in such a scaling, we have not rescaled $x_4$ or $x_5$.  Therefore, when applying this symmetry to the space of solutions, we know that $u_0\rightarrow C\times u_0$ however, $L_4$ and $L_5$ do not scale at all (as they are related to the asymptotic separation of $x_4$ and $x_5$).  This further accounts for the same behavior for the D5 embeddings found in \cite{Antonyan:2006pg}.

One can use such arguments to get the correct behavior in a host of situations.  Of particular interest is the situation where the metric is of the form $g_{i j}(u) dx^i dx^j+f(u)\times(du^2+u^2d\Omega_M^2)$ where all metric components along the $i,j$ directions are functions of $u$ only, and $d\Omega_M^2$ is the metric of some compact space spanned by coordinates $\theta^i$.  We further assume that the dilaton is a function only of $u$.  The reduced one dimensional Lagrangian with $x^i=\xi^i$ and $\theta^i=\theta^i(r)$ will always be of the form $G(u)\sqrt{\left(1+u^2d\Omega_M^2\right)}$ where now all of the $d$s in the metric $d\Omega_M^2$ are to be read as derivatives with respect to $u$.  In such a situation, $u\rightarrow C\times u$ is a Lie point symmetry of the action if $G(C\times u)=G(u)\times C^{\Delta}$, i.e. $G(u)=u^{\Delta}$.  This covers most extremal brane type embeddings.  Further, such behavior is intuitive for ``flat'' type embeddings of the probe brane in supersymmetric situations.  For a supersymmetric situation, the probe brane does not ``see'' the background it is in, and simply goes along a straight path (straight in the sense of the flat geometry with no backreacted branes).  Therefore, the angular separation is always a ``north pole/south pole'' type (at least asymptotically), and will not depend on $u_0$ at all.  As an example, see the discussion of \cite{Karch:2002sh} (equation (8)).  Of course the angular separation may be changed by some constant if there is a conical singularity at the origin of the transverse space (at $u=0$), as in the more recent calculation \cite{Kuperstein:2008cq} (see the last paragraph of section 3), however there should still be no dependence on $u_0$.

We can explore such possibilities in general.  Consider a situation where the metric is of the form $C_1r^n dx^i dx^j \eta_{i j} + C_2r^m dr^2+ C_3r^{m+2} d\Omega^2$, and the dilaton is of the form $\exp(\phi)\propto r^q$.  If one reduces the the DBI action to a one dimensional action via $x^i=x^i(r), \theta^i=\theta^i(r)$ (with an appropriate number of static gauge coordinates), one will always have the Lie point symmetry $r\rightarrow C\times r$, $x^i\rightarrow C^{\frac{m-n+2}{2}}x^i$, $\theta^i\rightarrow \theta^i$.  In such a situation, the asymptotic separation in the $x^i$ is $\Delta x^i\propto u_0^{\frac{m-n+2}{2}}$, and the angular separation $\Delta \theta^i={\rm constant}$.  For example in the case of the extremal Sakai-Sugimoto model \cite{Sakai:2004cn,Sakai:2005yt} $n=3/2,m=-3/2$ and so $ \Delta x^4\propto u_0^{-\frac{1}{2}}R^{\frac{3}{2}}$, where we have used $R$ to make the correct units.  This answer is indeed correct \cite{McNees:2008km} (see equation (2.18) of this work).

Our situation is somewhat special, where both the sphere and the flat space component have the same warp factor up to a constant.  This is the reason that in our extremal case, $L_4$ and $L_5$ do not depend at all on $u_0$.  It is a statement that a pure rescaling in $u$ is a Lie point symmetry, without requiring extra rescalings of the functions $x^i$.  Further, in such situations, the boundary conditions are not affected by varying $u_0$.  Therefore, one may expect such a mode to be normalizable, and correspond to a massless mode.  One usually thinks of such a mode as a  Goldstone boson of some spontaneously broken symmetry.  Here, the symmetry is the Lie point symmetry promoted to a full Noether symmetry, as discussed in the footnote earlier.  This is a full (non linear) symmetry of the DBI action.  It would be interesting to see what the symmetry of the DBI action corresponds to on the field theory side.

We now turn back to the embeddings at hand.  Note that using the SO(2) symmetry of the background (mixing $x_4$ and $x_5$), one may rotate and translate to a frame where $x_4=0$, and further, $x_5(u)\in (-\frac{R\pi}{6}\cdots\frac{R\pi}{6})$.  Such a frame change will mix any periodic identifications one has made in $x_4$ and $x_5$ (by a slant identification ``$\tau$''), however, the local physics of the embedding will not be dependent on such considerations.  Therefore, the point $x_5=0$ corresponds to where the embedding has reached $u_0$.  Therefore, we may summarize the embedding as
\be
u(x_5)=\left(\frac{p_5}{\sin\left(\frac{3 \left(x_5+\frac{R\pi}{6}\right)}{R}\right)}\right)^{\frac13}.
\ee
This is suggestive of a change of variables.  We make the change
\bea
y&=&u^3\sin\left(\frac{3 \left(x_5+\frac{R\pi}{6}\right)}{R}\right) \\
v&=&u^3\cos\left(\frac{3 \left(x_5+\frac{R\pi}{6}\right)}{R}\right).
\eea
In these coordinates, the metric becomes
\bea
ds^2&=&\frac{\left(y^2+v^2\right)^{\frac16}}{R}\left(\eta_{\mu \nu} dx^{\mu} dx^{\nu}+dx_4^2\right) + \frac{R}{9\left(y^2+v^2\right)^{\frac56}}\left(dy^2+dv^2\right) \nn \\
&&\qquad \qquad \qquad \qquad+R\left(y^2+v^2\right)^{\frac16}d\Omega_3^2.
\eea
We change to ``7 brane frame'' via
\be
G_{I J}= \exp\left(-\phi \frac{2}{8}\right) g_{I J}.
\ee
Changing frame in this way gives that $\sqrt{\det G_p}= \sqrt{\det g_p} \exp(-\phi)$ where $G_p$ is the pullback of the above $G$ to the D7 world volume.  In this frame
\bea
\frac{g_s^{\frac14}}{R^{\frac14}}ds^2_G&=&\frac{\left(y^2+v^2\right)^{\frac18}}{R}\left(\eta_{\mu \nu} dx^{\mu} dx^{\nu}+dx_4^2\right) + \frac{R}{9\left(y^2+v^2\right)^{\frac78}}\left(dy^2+dv^2\right) \nn \\
&& \qquad \qquad \qquad \qquad +R\left(y^2+v^2\right)^{\frac18}d\Omega_3^2.
\eea
The embedding of the D7 is now given by $y=p_5=u_0^3, x_4=0$.  This gives the pullback metric to be
\be
\frac{g_s^{\frac14}}{R^{\frac14}}ds^2_G=\frac{\left(u_0^6+v^2\right)^{\frac18}}{R}\left(\eta_{\mu \nu} dx^{\mu} dx^{\nu}\right) + \frac{R}{9\left(u_0^6+v^2\right)^{\frac78}}\left(dv^2\right)+R\left(u_0^6+v^2\right)^{\frac18}d\Omega_3^2.
\ee
One can now see why $y=u_0^3$ is a good embedding for any $u_0$: the determinant of the above metric does not depend on the warp factors $(u_0^6+v^2)$!  We will see the solution associated with shifting $u_0$ in the linearized equations of motion, which we will study in the next section.  We now add a word of caution.  The above symmetry scaling $u_0$ to different values is only a classical symmetry.  This may not be true quantum mechanically, in analogy with the expectations for $u_\Lambda$ expressed in section 1 \cite{Aharony:2004xn}.

\subsection{Scalar Mesons}

We now turn to the fluctuations of the above embedding.  To do so, we expand the fields as $y=u_0^3+w(v,x^{\mu}),x_4=0+x_4(v,x^{\mu})$.  Therefore, we find
\bea
&&\frac{g_s^{\frac14}}{R^{\frac14}}ds^2_G= \frac{\left((u_0^3+w)^2+v^2\right)^{\frac18}}{R}\left(dx^{\mu} dx_{\mu}\right) + \frac{R}{9\left((u_0^3+w)^2+v^2\right)^{\frac78}}\left(dv^2\right) \nn \\
&& \qquad \qquad \qquad \qquad \qquad +R\left((u_0^3+w)^2+v^2\right)^{\frac18}d\Omega_3^2  \\
&&+\frac{R}{9\left(u_0^6+v^2\right)^{\frac78}}\left(\pa_v w dv+\pa_\mu w dx^\mu\right)^2+\frac{\left(u_0^6+v^2\right)^{\frac18}}{R}\left(\pa_v x_4 dv+\pa_\mu x_4 dx^\mu \right)^2.\nn
\eea
We wish to expand $\det(G_p)$ to second order in $w$ and $x_4$.  With the above parametrization, this is now easy, as we may think of the third line as a perturbation of a metric given by the first line.  Therefore, we define
\bea
&& \frac{g_s^{\frac14}}{R^{\frac14}}ds_0^2=\frac{\left((u_0^3+w)^2+v^2\right)^{\frac18}}{R}\left(dx^{\mu} dx_{\mu}\right) + \frac{R}{9\left((u_0^3+w)^2+v^2\right)^{\frac78}}\left(dv^2\right) \nn \\
&& \qquad \qquad \qquad \qquad \qquad +R\left((u_0^3+w)^2+v^2\right)^{\frac18}d\Omega_3^2
\eea
and use this to define a metric $G_p$.  Now, to leading order, the expanded metric
\be
\det(G_p+h)=\det(G_p)\left(1+G_p^{a_1 a_2} h_{a_1 a_2}\right).
\ee
In the second term of the above, we may substitute $w=0$ into the metric $g_p$ as $h$ is already second order in $w$ and $x_4$.  Further, $\det(g_p)$ does not depend on $w$ or $x_4$, and so there is no need to expand further.  We therefore find that, to second order
\bea
\int d\Omega_3\sqrt{\det(G_p)}=&&\Omega_3\frac{R}{3 g_s}\Bigg(1+ \frac12\pa_\mu x_4\pa^{\mu}x_4+\frac{9(u_0^6+v^2)}{R^2}\frac12(\pa_v x_4)^2 \nn \\ &&
\qquad \qquad + \frac{R^2}{9(u_0^6+v^2)}\frac12 \pa_{\mu} w \pa^{\mu} w+\frac12(\pa_v w)^2\Bigg).
\eea
The second order action we wish to consider is therefore
\bea
&&S=-\frac{\kappa_7 R \Omega_3}{g_s}\int dv d^4x\Bigg(\frac{\zeta^{-1}}{9}\frac12 \pa_{\mu} w \pa^{\mu} w+\frac12(\pa_v w)^2+\frac12\pa_\mu x_4\pa^{\mu}x_4+9\zeta\frac12(\pa_v x_4)^2\Bigg) \nn
\eea
with
\be
\zeta\equiv \frac{u_0^6+v^2}{R^2}.
\ee
The above action has positive definite Hamiltonian, so we expect the embedding to be stable.

The eigenvalue problem we wish to satisfy is the following
\bea
\pa_V \left(9(1+V^2)\pa_V \psi_{x,M_x}\right)&=& -M_x^2\psi_{x,M_x}(V) \\
9(1+V^2)\pa_V^2 \psi_{w,M_w}(V)&=&-M_w^2 \psi_{w,M_w}(V).
\eea
where the 4D masses are given by $m_i=\frac{M_i}{R}$ and we have defined $V=\frac{v}{u_0^3}$.  These have the general solutions
\bea
\psi_{x,M_x}(V)&=& C_{1,x}P\left(\frac{\sqrt{9-4M_x^2}}{6}-\frac12;-iV\right)+  \nn \\ &&+\qquad C_{2,x}Q\left(\frac{\sqrt{9-4M_x^2}}{6}-\frac12;-iV\right)\\
\psi_{w,M_w}(V)&=&C_{1,w}(1+V^2)\;_2F_1\left(\frac34+\frac{\sqrt{9-4M_w^2}}{12},\frac34-\frac{\sqrt{9-4M_w^2}}{12} ;\frac12;-V^2 \right)  \\
&&+C_{2,w}V(1+V^2)\;_2F_1\left(\frac54+\frac{\sqrt{9-4M_w^2}}{12},\frac54-\frac{\sqrt{9-4M_w^2}}{12};\frac32;-V^2 \right)\nn
\eea
where $P$ and $Q$ are Legendre functions, and $\;_2F_1$ is the hypergeometric function.  There are possible bound states in the regime
\be
0\leq m_{x,w}=\frac{M_{x,w}}{R}\leq\frac32 \frac{1}{R}.
\ee
Curiously, we will find this to be true in the next section as well, where we consider an embedding into the near extremal background.

We now change the eigenvalue problem to Sch\"odinger form to analyze the above differential equations.  We do so by changing coordinates and variables
\bea
V&=&\sinh(K) \nn \\
\psi_{w,M_w}(K)&=&\sqrt{\cosh(K)}\Psi_{w,M_w}(K) \nn \\
\psi_{x,M_x}(K)&=&\frac{1}{\sqrt{\cosh(K)}}\Psi_{x,M_x}(K).
\eea
This brings the equations to the following form
\bea
&&\pa_K^2 \Psi_{w,M_w}(K)-\left(\frac14-\frac34\frac{1}{\cosh(K)^2}\right)\Psi_{w,M_w}(K)=-\frac19M_w^2\Psi_{w,M_w}(K) \\
&&\pa_K^2 \Psi_{x,M_x}(K)-\left(\frac14+\frac14\frac{1}{\cosh(K)^2}\right)\Psi_{x,M_x}(K)=-\frac19M_x^2\Psi_{x,M_x}(K).
\eea
The second line clearly has no normalizable bound states, where as the first at least has the possibility.  In fact, the above set of differential equations is {\it exactly} the same as those encountered in the next section, with a simple modification.

In the next section we show that there is only one bound state for the equation of the first type above, and  in fact this is the the $M_w=0$ mode anticipated by the discussion above.  For $M_w=0$, we find
\be
\Psi_{w,M_w=0}(K)=\frac{C_{1,w}}{\sqrt{\cosh(K)}}+\frac{C_{2,w}\sinh(K)}{\sqrt{\cosh(K)}}.
\ee
The first is clearly normalizable, where the second is clearly non normalizable.  Further, taking the normalizable solution, we find that the profile in $\psi_{w,M_w=0}=C_{1,w}$, exactly corresponding to shifts in the definition of $u_0$.  We will fully analyze the above type of differential equations in the next section.

\subsection{Vector Mesons}

We now analyze the fluctuations of the world volume vector gauge field.
As in the previous section, we find that
\bea
&&=-\kappa_7\int d\Omega_3 \int dx^\mu dv e^{-\phi} \sqrt{g_p} \left(1+\frac14 F_{a_1 b_1} F_{a_2 b_2} g_p^{a_1 a_2}g_p^{b_1 b_2}+\cdots \right).
\eea
which for this case gives
\be
S= -\kappa_7\Omega_3\int dx^\mu dv \frac{R}{g_s} \left(1+\frac14 \frac{g_s^{\frac12}R^{\frac32}}{\left(u_0^6+v^2\right)^{\frac14}}F_{\mu \nu}F^{\mu \nu}+\frac12 \frac{9 g_s^{\frac12}\left(u_0^6+v^2\right)^\frac34}{R^{\frac12}}F_{\mu v}F^\mu{}_v+\cdots \right)
\ee
where we now use $\eta^{\mu \nu}$ to raise and lower indices.
The equations of motion read
\bea
R^2 \pa^\mu F_{\mu \nu}+\left(u_0^6+v^2\right)^\frac14\pa_v \left(\left(u_0^6+v^2\right)^\frac34F_{v \nu}\right)=0 \\
\pa^\mu F_{\mu v}=0.
\eea
As before we expand
\be
A_\mu=\sum_n B^{(n)}_\mu(x) \psi_n(v), \quad A_v=\sum_n \pi_n(x) \phi_n(v)
\ee
and gauge fix $\pa^\mu B_\mu=0$ to find
\bea
\sum_n R^2 \pa^\mu\pa_\mu B^{(n)}_\nu \psi_n + \left(u_0^6+v^2\right)^{\frac14}\pa_v \left(\left(u_0^6+v^2\right)^{\frac34}\pa_v \psi_n\right)B^{(n)}_\nu =0 \\
\pa_v\left(\left(u_0^6+v^2\right)^{\frac34} \pa_\nu \pi_0 \phi_0\right)=0\rightarrow \phi_0\propto \frac{1}{\left(u_0^6+v^2\right)^{\frac34}}.
\eea
and for $n\neq 0$, $\phi_n=0$.

We may change the $\psi_n$ equations to Schr\"odinger form via
\bea
&& \frac{v}{u_0^3}=\sinh(K) \nn \\
&& \psi_{(n)}(K)=\frac{1}{\left(\cosh(K)\right)^{\frac14}}\Psi_n
\eea
and taking $\pa^\mu \pa_\mu B^{(n)}=m_n^2B^{(n)}\equiv \frac{M_n^2}{R^2}B^{(n)}$ we find
\be
\pa_K^2 \Psi(K) -\frac{1}{144} \Psi(K) \left(9+\frac{27}{\cosh(K)^2}\right)=-\frac19 M_n^2 \Psi(K).
\ee
This obviously has no normalizable modes.  This means that the only 4D meson from this sector is the massless pion $\pi_0$.  We will see the same phenomena in the next section.

\section{A simple antipodal embedding}

In this section we will be concerned with the antipodal embedding $P_4=0,P_5=0$.  This is the rare exception when we have an analytic solution that comes into the region $u\sim u_\Lambda$.  We will find it convenient to start with the $z$ coordinates introduced earlier
\be
u^2=u_\Lambda^2+z^2.
\ee
In these variables, the metric reads
\bea
ds^2&&=\frac{\sqrt{u_\Lambda^2+z^2}}{R}\left(\eta_{\mu \nu} dx^{\mu}dx^{\nu}+dx_4^2\right)+R\sqrt{u_\Lambda^2+z^2} d\Omega_3^2 \nn \\
&&\qquad \qquad +\frac{R}{\sqrt{u_\Lambda^2+z^2}}\left(dz^2+z^2\frac{dx_5^2}{R^2}\right)
\eea
and we make the following coordinate change
\bea
w\equiv z \sin\left(\frac{x_5}{R}\right) \\
v\equiv z \cos\left(\frac{x_5}{R}\right)
\eea
so that
\bea
ds^2&&=\frac{\sqrt{u_\Lambda^2+z^2}}{R}\left(\eta_{\mu \nu} dx^{\mu}dx^{\nu}+dx_4^2\right)+R\sqrt{u_\Lambda^2+z^2} d\Omega_3^2 \nn \\
&&\qquad \qquad +\frac{R}{\sqrt{u_\Lambda^2+z^2}}\left(dw^2+dv^2\right).
\eea
Now, the solution to the embedding reads $x_4(v)=w(v)=0$ (we will use $v$ as the world volume coordinate).

\subsection{Scalar mesons}

We now expand the action to quadratic order in $x_4(v,x^\mu), w(v,x^{\mu})$.  This is most easily accomplished by going to a ``seven brane'' frame
\be
G_{I J}= \exp\left(-\phi \frac{2}{8}\right) g_{I J}.
\ee
Changing frame in this way gives that $\sqrt{\det G_p}= \sqrt{\det g_p} \exp(-\phi)$ where $G_p$ is the pullback of the above $G$ to the D7 world volume.  We first expand the modified line element to second order in fluctuating fields $x_4$ and $w$, and we find
\bea
g_s^{\frac14} ds_G^2&& =\zeta^{\frac 38}\eta_{\mu \nu} dx^\mu dx^\nu + R^2\zeta^{\frac38}d\Omega_3^2+\zeta^{-\frac58} dv^2 \nn \\
&&+\zeta^{\frac38}\left(\pa_v x_4 dv+\pa_\mu x_4 dx^\mu\right)^2+\zeta^{-\frac58}\left(\pa_v w dv+ \pa_\mu w dx^\mu\right)^2 \nn \\
&&+\frac38 \zeta^{-\frac58}\frac{w^2}{R^2}\eta_{\mu \nu}dx^\mu dx^\nu+R^2\frac38\zeta^{-\frac58}\frac{w^2}{R^2}d\Omega_3^2 -\frac58\zeta^{-\frac{13}{8}}\frac{w^2}{R^2}dv^2+\cdots
\eea
with
\be
\zeta\equiv \frac{u_\Lambda^2+v^2}{R^2}.
\ee
Note that the $\zeta$ defined in this section is different than the $\zeta$ defined in the last section.

Next, we expand the action to second order and find
\bea
\int d^4x^\mu dv d\Omega_3 {\sqrt{-G_p}}&&= \Omega_3 \int d^4x^\mu dv \zeta\Bigg[1+\frac12 \pa_\mu x_4 \pa^\mu x_4+\frac12 \zeta (\pa_v x_4)^2  \nn \\
&&+ \frac12 \zeta^{-1}\pa_\mu w \pa^\mu w+\frac12 (\pa_v w)^2+\frac12 \frac{2}{R^2} w^2\Bigg]
\eea
which has positive definite hamiltonian, and so we expect that this embedding is stable.

The proper eigenvalue problem is
\bea
\zeta^{-1}\pa_v \left(\zeta^2 \pa_v \psi_{x,m_x}(v)\right)=-\frac{M_x^2}{R^2} \psi_{x,m_x}(v) \nn \\
\pa_v \left(\zeta \pa_v \psi_{w,m_w}(v)\right)=\left(\frac{2}{R^2}-\frac{M_w^2}{R^2}\right)\psi_{w,m_w}
\eea
where the 4D masses are $m_x=\frac{M_x}{R}$ and $m_w=\frac{M_w}{R}$.
The solution to these are the following
\bea
\psi_{x,M_x}(V)&&=C_{x,1}\;_2F_1\left(\frac34-\frac{\sqrt{9-4M_x^2}}{4},\frac34+\frac{\sqrt{9-4M_x^2}}{4};\frac12;-V^2\right) \nn \\
&& \quad+  C_{x,2} V \;_2F_1\left(\frac54-\frac{\sqrt{9-4M_x^2}}{4},\frac54+\frac{\sqrt{9-4M_x^2}}{4};\frac32;-V^2\right) \\
\psi_{w,M_w}(V)&&=C_{w,1}P\left(\frac{\sqrt{9-4M_w^2}}{2}-\frac12;-iV\right) \nn \\ &&\quad +C_{w,2}Q\left(\frac{\sqrt{9-4M_w^2}}{2}-\frac12;-iV\right)
\eea
where
\be
V=\frac{v}{u_\Lambda}
\ee
and $\;_2F_1$ is the hypergeometric series, and $P$ and $Q$ are Legendre functions.  Note again that the definition of the above $V$ differs from the last section.

There are possible bound states in the range
\be
0 \leq m_{x,w}=\frac{M_{x,w}}{R}\leq \frac32 \frac{1}{R}.
\ee
If $m_x$ or $m_w$ exceed this range of values, the indices of the above functions become complex.  One can show that around $V=\infty$ these go as appropriately normalized plane wave solutions, and so represent a continuum of states.  This corresponds to a mass scale at which the scalar mesons may no longer be considered 4 dimensional.

We expect that the indices of the hypergeometric equation being integers is special, and we find that in the case $M_x=\sqrt{2}$ that
\be
\psi_{x,M_x=\sqrt{2}}(V)=\frac{C_{x,1}+C_{x,2}V}{V^2+1}.
\ee
In such a case we see that $C_{x,2}=0$ is a normalizable solution.  We will now give an argument that this is the only bound state of the above eigenvalue equations.

\begin{figure}[ht]
\centering
\includegraphics[width=.4\textwidth]{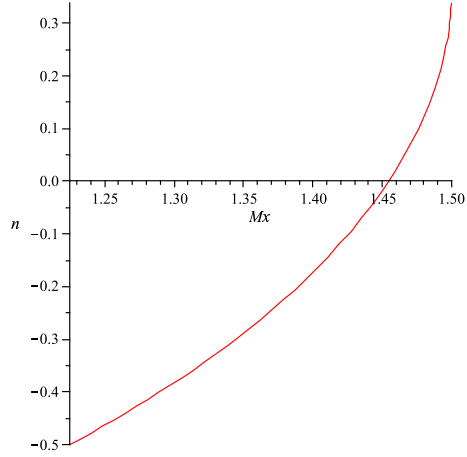}
\caption{Plot of $n(M_x)$.  Note that the only integer solution is $n=0$, and that this approximates the known solution of $M_x=\sqrt{2}$.}
\label{integern}
\end{figure}

We will change the one dimensional problem in hand to Schr\"odinger form.  To do so, we take the change of variables and functions
\bea
V=\sinh(K)\\
\psi_{w,M_w}(K)=\frac{\Psi_{w,M_w}(K)}{\sqrt{\cosh(K)}}  \\
\psi_{x,M_x}(K)=\frac{\Psi_{x,M_x}(K)}{\left(\cosh(K)\right)^{\frac32}}.
\eea
Then, the equations above become
\bea
\frac{\pa^2}{\pa K^2} \Psi_{w,M_w}(K)-\left(\frac{9}{4}+\frac{1}{4}\frac{1}{\cosh(K)^2}\right)\Psi_{w,M_w}(K)=-M_w^2\Psi_{w,M_w}(K) \\
\frac{\pa^2}{\pa K^2} \Psi_{x,M_x}(K)-\left(\frac{9}{4}-\frac{3}{4}\frac{1}{\cosh(K)^2}\right) \Psi_{x,M_x}(K)=-M_x^2 \Psi_{x,M_x}(K).
\eea
In the above problem, we are looking for bound states whose energy is interpreted as $E=M_i^2$.  Clearly there are no bound states for the potential of the first kind (it is a hill, rather than a well).  In the second case, there is at least the possibility that there are bound states.  Using the WKB approximation, we take
\be
n=-\frac12+\frac{1}{\pi}\int_{-\cosh^{-1}\left(\frac{3}{\sqrt{-12 M_x^2+27}}\right)}^{\cosh^{-1}\left(\frac{3}{\sqrt{-12 M_x^2+27}}\right)}dK\sqrt{M_x^2-\frac{9}{4}+\frac{3}{4}\frac{1}{\cosh(K)^2}}.
\ee
There is a bound state when $n$ is an integer.  Clearly $\sqrt{\frac32}<M_x<\sqrt{\frac94}$ by considering the maximum and minimum of the potential, and so we plot the above function in this range in figure \ref{integern}.  The plot of $n$ crosses an integer value only at $n=0$.  This crossing is close to the exact solution of $M_x=\sqrt{2}$, but not precise.  We may expect this low level of accuracy for the least energetic bound state, but generically, the accuracy of the WKB approximation is improved for higher energy bound states.  Therefore, we trust that there are no other bound state solutions.

One further note is in order.  For the above configuration, there is no scaling symmetry of the DBI action (as there was in the last section).  Therefore, it seems plausible that the above mode is a pseudo goldstone mode associated with the scaling symmetry.  Because the symmetry is not exact (due to the function $f$), the mass is lifted from 0.  It would be interesting to find a way to track the mode from small to large values of $u_0/u_\Lambda$ and see if this is indeed the case.  This would further track from being a pure $w$ excitation (when $u_0\gg u_\Lambda$, as in the last section) to being a pure $x_4$ excitation (for $u_0=u_\Lambda$, in this section).  Finding the correct linear combination that was a bound 4D state, and diagonalized the mass matrix in the general case would be instructive.

\subsection{Vector mesons}

As in the previous section, we find that
\bea
&&=-\kappa_7\int d\Omega_3 \int dx^\mu dv e^{-\phi} \sqrt{g_p} \left(1+\frac14 F_{a_1 b_1} F_{a_2 b_2} g_p^{a_1 a_2}g_p^{b_1 b_2}+\cdots \right).
\eea
which for this case gives
\bea
S_{F^2}&&= -\kappa_7\Omega_3\int dx^\mu dv \sqrt{\frac{u_\Lambda^2 + v^2}{R^2}}^2 \frac{R^3}{4g_s} \left(\sqrt{\frac{u_\Lambda^2 + v^2}{R^2}}^{-2}F_{\mu \nu}F^{\mu \nu}+2F_{\mu v}F^\mu{}_v\right) \nn \\
&&=-\frac{\kappa_7\Omega_3R^3}{g_s}\int dx^\mu dv \frac{1}{4} \left(F_{\mu \nu}F^{\mu \nu}+2\sqrt{\frac{u_\Lambda^2 + v^2}{R^2}}^2F_{\mu v}F^\mu{}_v\right).
\eea
The equations of motion read
\bea
\pa_\mu F^{\mu \nu}+\pa_v F^{v \nu}=0 &\leftrightarrow & \pa^\mu F_{\mu \nu}+\pa_v \left(\frac{u_\Lambda^2+v^2}{R^2}F_{v \nu}\right)=0 \\
\pa_\mu F^{\mu v}=0 &\leftrightarrow & \pa^\mu F_{\mu v}=0.
\eea
As before we expand
\be
A_\mu=\sum_n B^{(n)}_\mu(x) \psi_n(v), \quad A_v=\sum_n \pi_n(x) \phi_n(v)
\ee
and gauge fix $\pa^\mu B_\mu=0$ to find
\bea
\sum_n \pa^\mu\pa_\mu B^{(n)}_\nu \psi_n + \pa_v \left(\frac{u_\Lambda^2+v^2}{R^2}\pa_v \psi_n\right)B^{(n)}_\nu =0 \\
\pa_v\left(\frac{u_\Lambda^2+v^2}{R^2} \pa_\nu \pi_0 \phi_0\right)=0\rightarrow \phi_0\propto \frac{1}{1+\frac{v^2}{u_\Lambda^2}}.
\eea
In this case, the eigenvalue problem is
\be
\pa_V \left((1+V^2)\pa_V \psi_n(V)\right)=-M_{A,n}^2\psi_n(V).
\ee
where again $V=\frac{v}{u_\Lambda}$.

We change to Schr\"odinger form via
\bea
V&&=\sinh(K) \\
\psi_n(K)&&=\frac{1}{\sqrt{\cosh(K)}}\Psi_n(K)
\eea
and find
\be
\left(\pa_K^2 -\frac14\left(1+\frac{1}{\cosh(K)^2}\right)\right)\Psi_n(K)=-M_{A,n}^2\Psi_n(K)
\ee
which obviously has no normalizable states.  Note that the eigenvalue problem is actually the same as the $M_w$ problem above (with a shift in the potential).  Hence, the only 4D meson from this sector is again the massless pion $\pi_0$.

\subsection{Baryons}

We now turn to the question of determining the size of the baryon, and for concreteness, we will model this by an instanton in the $U(N)$ gauge field living on the brane.  The size of the baryon will be stabilized by the presence of the CS term as in the SS model \cite{Sakai:2004cn,Hata:2007mb}.  We will only use the above quadratic action plus the CS term arising from the term
\be
S_{CS}=\mp \kappa_7 \int \Tr\left(\exp\left(F\right)\right)\wedge C_2
\ee
where $C_2$ is the potential leading to $F_3$ of the background, and further $F=dA-i[A,A]$ is the U($N_f$)field strength.  The appropriate term is the $F^3$ term.  We rescale $F$ by the canonical factor of $2 \pi \alpha'$, and integrate by parts to find that
\be
S_{CS}=\mp \kappa_7 \frac{1}{3!}(2\pi \alpha')^3\int \omega_5(A)\wedge F_3=\mp \frac{N_c}{24}\int \omega_5(A)
\ee
where
\be
\omega_5(A)=\rm{\Tr}\left(AF^2-\frac{i}{2}A^3F-\frac{1}{10} A^5\right)
\ee
and the integration is now five dimensional.  This is exactly the term found in \cite{Hata:2007mb}.  We now turn to the quadratic action
\be
S_{F^2}=-\frac{\kappa_7\Omega_3R^3}{g_s}\int dx^\mu dv \frac{1}{2}{\rm Tr}\left(F_{\mu \nu}F^{\mu \nu}+2\sqrt{\frac{u_\Lambda^2 + v^2}{R^2}}^2F_{\mu v}F^\mu{}_v\right)
\ee
and perform the coordinate transformation $x^\mu=R\hat{x}^\mu$, $v=u_\Lambda V$, and further scale $F$ by $2\pi \alpha'$ to find
\bea
S_{F^2}&=&-\frac{\kappa_7\Omega_3R^3u_\Lambda(2\pi \alpha')^2}{g_s}\int d\hat{x}^\mu dV \frac{1}{2}{\rm Tr} \left(F_{\mu \nu}F^{\mu \nu}+2\left(1 + V^2\right)F_{\mu V}F^\mu{}_V\right) \nn \\
&=& -\frac{u_{\Lambda}}{2(2\pi)^3 \sqrt{\alpha'}} \frac{(g_s N_c)^{\frac32}}{g_s}\int d\hat{x}^\mu dV \frac{1}{2}{\rm Tr} \left(F_{\mu \nu}F^{\mu \nu}+2\left(1 + V^2\right)F_{\mu V}F^\mu{}_V\right) \nn \\
\eea
where $\mu,\nu$ indices are still raised and lowered by $\eta_{\mu \nu}$.  We will drop the $\hat{\;}$ in the following, knowing that $V$ is measured in $u_\Lambda$ units and $x^\mu$ is measured in $R$ units.  Note that in the limit that $V\rightarrow 0$ that the above action goes to that of the free 5D Yang-Mills (YM) theory, as it does in \cite{Hata:2007mb}.  We therefore write down the combined action
\bea
S=S_{F^2}+S_{CS}.
\eea

Next, we wish to consider the mass of the baryon to leading order in a ``small size'' expansion.  We will do this for the SU(2) BPST instanton by checking the energy as a function of the instanton size.  The BPST instanton configuration is
\bea
&&\kern-1em F_{ij}=\frac{2\rho^2}{(\xi^2+\rho^2)^2}\epsilon_{ija}\tau^a,\quad F_{iV}=\frac{2\rho^2}{(\xi^2+\rho^2)^2}\tau_i,
\quad \xi=\sqrt{(\vec{x}-\vec{x}_0)^2+(V-V_0)^2}.
\eea
The constants $(\vec{x}_0,V_0)$ parameterize the location and $\rho$ parameterizes the size.  Above, $\tau^i$ are the Pauli matrices that satisfy ${\rm Tr}(\tau^a \tau^b)=2\delta^{ab}$.  We plug this into the Lagrangian density, and integrate to give the energy of this configuration (measured in units of $R$, as we must ``unscale'' time to get something with units)
\bea
E(\rho,V_0) R &=&\frac{u_{\Lambda} }{2(2\pi)^3 \sqrt{\alpha'}} \frac{(g_s N_c)^{\frac32}}{g_s}\int d^3\hat{x} dV \frac{1}{2}{\rm Tr} \left(F_{ij}F^{ij}+2\left(1 + V^2\right)F_{i V}F^i{}_V\right) \nn \\
&=& \frac{u_{\Lambda} }{2(2\pi)^3 \sqrt{\alpha'}} \frac{(g_s N_c)^{\frac32}}{g_s}\int dV \frac{3\rho^4\pi^2\left(2+V^2\right)}{\left((V-V_0)^2+\rho^2\right)^{\frac52}} \nn \\
&=& \frac{u_{\Lambda}}{2(2\pi)^3 \sqrt{\alpha'}} \frac{(g_s N_c)^{\frac32}}{g_s}(4\pi^2 V_0^2+2\pi^2\rho^2+8\pi^2)
\eea
which is only minimized at $V_0=0$ and $\rho=0$.

Note that the above energy is the mass of the baryon in this model, which, plugging in $R=\sqrt{N_c \alpha' g_s}$ above gives
\be
E(0,0)=M_b=\frac{N_c u_\Lambda}{2\pi\alpha'}.
\ee
This is exactly reproduced by the energy of a D3 wrapping the S$^3$ (sitting at $u=u_\Lambda$) in the geometry,
\bea
\kappa_3 \int e^{-\phi}\det(g_p)&=&(2\pi)^{-3}(\alpha')^2 \int dx^0  d\Omega_3 \frac{R}{u_\Lambda g_s}\sqrt{R^2 u_\Lambda^4} \nn \\
&=&(2\pi)^{-3}(\alpha')^{-2} 4\pi^2 \frac{R^2}{g_s} u_\Lambda\int dx^0=\frac{N_c u_\Lambda}{2\pi \alpha'} \int dx^0.
\eea

Before we address the effect of the CS term on the size of the baryon, we first wish to identify a perturbative paremeter that will make the $F^2$ term parametrically larger than the CS term.  To do so, we note that the coefficient of the $F^2$ term as
\be
S_{F^2}=\frac{(u_\Lambda^2 R^2)^{\frac12}}{\alpha'} \frac{24}{2(2\pi)^3} \frac{N_c}{24}\int d\hat{x}^\mu dV \frac{1}{2}{\rm Tr} \left(F_{\mu \nu}F^{\mu \nu}+2\left(1 + V^2\right)F_{\mu V}F^\mu{}_V\right) \nn \\
\ee
and so the relative coefficient is $\frac{(u_\Lambda^2 R^2)^{\frac12}}{\alpha'} \frac{24}{2(2\pi)^3}\equiv \hat{\lambda}$.  We use this as the large parameter that we will be expanding in.  We expect corrections in the size of the instanton to be of order $\lambda^{-\frac12}$.  With this in mind we scale the spatial dimensions by
\be
\xh^i=\lh^{-\frac12} \xt^i\;i\in\{1\cdots3\}, V= \lh^{-\frac12}\vt.
\ee
The discussion now exactly mirrors that of the discussion in \cite{Hata:2007mb}, with the replacement in their formulas of $a=1/24$.  Further, the parameter $Z$ in their equations must match the appearance of $V_0$ in our calculation above.  Therefore, we simply use their result with the appropriate changes
\be
M_b=\frac{N_c u_\Lambda}{2\pi\alpha'}\left(1+\lh^{-1}\left(\frac{\vt_0^2}{2}+\frac{\tilde{\rho}^2}{4}+\frac{1}{320\pi^4 a^2}\frac{1}{\tilde{\rho}^2}\right) + {\mathcal O}\left(\lh^{-2}\right)\right)
\ee
where above we have emphasized the scaling with $\lh$ by including the tilde, i.e. $\tilde{V}_0=\lh^{\frac12}\times V_0$, and as we have mentioned before, $a=\frac{1}{24}$.  This gives that
\be
\tilde{\rho}^2=\frac{6}{\sqrt{5}\pi}\rightarrow \rho^2=\frac{6}{\sqrt{5}\pi}\frac{1}{\lh}
\ee
To restore units, we note that one must include a factor of $R$ in the above to get the order of the radius in the $x^i$ directions.  This gives a final result of
\be
\rho_x^2=\frac{6}{\sqrt{5}\pi}\frac{1}{\lh}R^2=\frac{4}{\sqrt{5}}\pi^2\frac{R}{u_\Lambda}\alpha'.
\ee
Physical length in this background near the point $u=u_\Lambda$ is measured in $dx^2/(R/u_\Lambda)$ units.  Therefore, the physical size of the above distribution is $\frac{2\pi}{5^{1/4}}\sim 4.2$ in string units.  This is not very big, and so string corrections may be needed.

\section{Deconfined phase.}

One may also wish to study the physics of the phase where the blackening factor $f$ appears on the timelike component of the metric.  For such a background, we have
\bea
\kern -1em ds^2&&=\frac{u}{R}\left(-f(u;u_T)dt^2+\delta_{ij} dx^{i}dx^{j}+dx_4^2+dx_5^2\right)+\frac{R}{u}\frac{du^2}{f(u;u_T)}+R u d\Omega_3^2
\eea
and
\be
\exp{\phi}=g_s \frac{u}{R}, \quad F_3=\frac{2 R^2}{g_s} \Omega_3.
\ee
In this case $u_T$ implies a horizon, however,classically  the temperature does not depend on this parameter at all (as the periodicity of $x_5$ in the last section).  Here
\be
\delta t=2\pi R= \frac{1}{T},
\ee
and further, we can take $u_T$ to be a measure of the energy density via
\be
E(u_T)=u_T^2\frac{V_5}{(2\pi)^5(\alpha')^4g_s^2}, \quad S=2\pi R E(u_T).
\ee
Above we have used the energy as measured for a near extremal D5 in flat ${\mathbb{R}}^{1,9}$, and $S$ is obtained via the Bekenstein-Hawking formula.

\begin{figure}[ht]
\centering
\includegraphics[width=.45\textwidth]{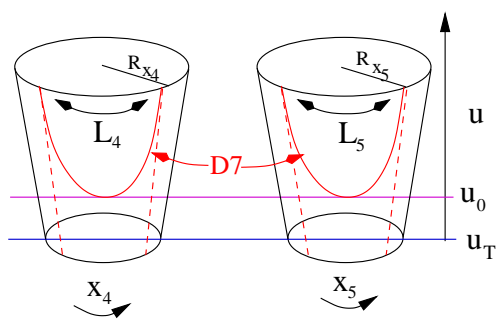}
\caption{The configuration of D7 branes in the near extremal D5 brane background with the thermal factor on the time component of the metric.  The dashed lines show a possible configuration with the branes falling into the horizon.}
\label{cigartherm}
\end{figure}

As in the last section, one can take the embedding of a D7 which is transverse to some combination of $u, x_4 x_5$.  Possible configurations are shown graphically in figure (\ref{cigartherm}).

Again, choosing $u$ to be the world volume coordinate of the brane $\rho$, we find the reduced action
\be
\frac{T_8}{g_s}V_4 V_3 \int d\rho \rho^3\sqrt{f(\rho;u_T)\left((\pa x_4)^2+(\pa x_5)^2\right)+\frac{R^2}{\rho^2}}.
\ee
As before, we find the conserved quantities
\bea
&&\frac{\rho^3 f(\rho; u_T)\pa x_4}{\sqrt{f(\rho;u_T)\left((\pa x_4)^2+(\pa x_5)^2\right)+\frac{R^2}{\rho^2}}}=P_4 u_T^3 \\
&&\frac{\rho^3 f(\rho; u_T)\pa x_5}{\sqrt{f(\rho;u_T)\left((\pa x_4)^2+(\pa x_5)^2\right)+\frac{R^2}{\rho^2}}}=P_5 u_T^3
\eea
which allows us to solve for $\pa x_4,\pa x_5$ as
\bea
&&\pah x_4=R\frac{P_4}{\sqrt{\rh^2f(\rh;1)\left(\rh^6f(\rh;1)-(P_4^2+P_5^2)\right)}} \\
&&\pah x_5=R\frac{P_5}{\sqrt{\rh^2f(\rh;1)\left(\rh^6f(\rh;1)-(P_4^2+P_5^2)\right)}}.
\eea
where in this section we define
\be
\rh=\frac{\rho}{u_T},\qquad \pah=\frac{\partial}{\partial \rh}
\ee
This allows us to immediately identify
\be
P_4^2+P_5^2\equiv P^2 \equiv \rh_0^6\left(1-\frac{1}{\rh_0^2}\right)
\ee
where $\rh_0$ is the minimal value to which the brane falls ($\rh_0=\rho_0/u_T=u_0/u_T$).  We note that the action is a function only of $\rh_0$.  The above configurations always give a connected brane antibrane for $P^2 >1$.

Further, we should note that because the functions $\frac{L_i(\rh_0)}{R}$ only depend on $\rh_0$, that for fixed boundary condition at infinity, varying $u_\Lambda$ does not affect $\rh_0$.  This is because to maintain the boundary conditions at infinity, one must vary $u_0$ in just such a way as to keep $u_0/u_T=\rh_0$ fixed.

One may now ask whether the connected configuration (considered above) is preferred or whether the configuration where the brane anti-brane pair simply fall into the horizon is preferred.  To measure this, we subtract the two (infinite) actions, considering the boundary conditions as being placed at some finite (but large) value of $u$, and then relaxing this position to infinity.  This defines our regularization procedure.

\begin{figure}[ht]
\centering
\includegraphics[width=.45\textwidth]{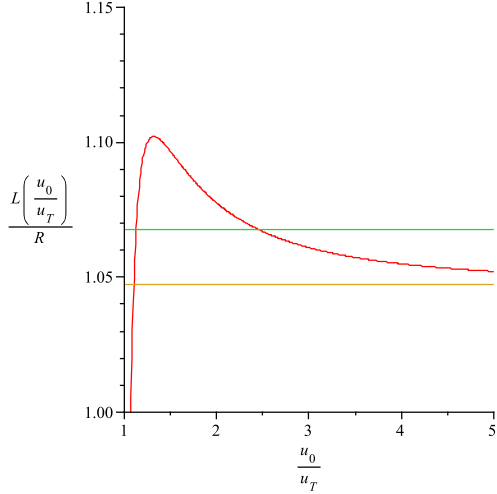}
\caption{$\frac{L\left(\rh_0=\frac{u_0}{u_T}\right)}{R}$ along with the critical value $\frac{L(\rh_0^c \approx 2.4456)}{R}\approx 1.068$, and the asymptotic value of $\frac{L(\rh_0=\infty)}{R}=\frac{\pi}{3}\approx1.047$, plotted as red green and yellow respectively.  Although we have not plotted it, the remainder of the graph for $\frac{L(\rh_0)}{R}$ goes to zero as $\rh_0\rightarrow 1$.}
\label{lofr0}
\end{figure}

The resulting difference we wish to consider is therefore
\bea
\frac{\Delta S}{2V_3 V_4 T_7 R u_T^3}&=&\int_{\rh_0}^\infty d\rh \rh^5 \sqrt{\frac{f(\rh;1)}{\rh^6f(\rh;1)-\rh_0^6f(\rh_0;1)}} -\int_1^\infty d\rh \rh^2 \nn \\
&=& \int_{\rh_0}^\infty d\rh \left(\rh^5 \sqrt{\frac{f(\rh;1)}{\rh^6f(\rh;1)-\rh_0^6f(\rh_0;1)}}-\rh^2\right) -\frac13(\rh_0^3-1).
\eea
We cannot evaluate this exactly, however, we may plot it numerically.
\begin{figure}[ht]
\centering
\includegraphics[width=.45\textwidth]{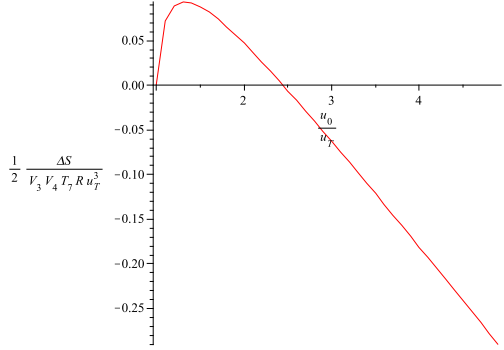}
\caption{The difference in actions between the ``U-shaped'' and ``$||$-shaped'' configurations plotted as a function of $\rh_0=\frac{u_0}{u_T}$.}
\label{DeltaS}
\end{figure}
The plot shows that for $\frac{u_0}{u_T}< \rh_0^c \approx 2.4456$ that the separated D7 $\overline{\rm D7}$ is preferred, and that for $\frac{u_0}{u_T}> \rh_0^c \approx 2.4456$ that the connected ``U-shaped'' configuration is preferred.  Because the temperature is not a function of $u_T$, we may not associate this with a phase transition, even in the space of energy density, of which $u_T$ is a measure.  To stress this point, we may easily plug this into the equations for $L_4$ and $L_5$ and find
\be
L(\rh_0^c)\equiv \sqrt{L_4(\rh_0^c)^2+L_5(\rh_0^c)^2}=1.068 R.
\ee
We note that in the above $R$ does not vary as we vary $u_T$, and neither does $L$ (as this is a boundary condition we are imposing).  We find therefore that the chiral symmetry breaking/restoration is purely a function of the boundary conditions imposed, and find that for $\frac{\pi}{3}R < L\lessapprox 1.068 R$ that chiral symmetry is present in the deconfining phase, and that for $L \gtrapprox 1.068 R$ chiral symmetry is broken in the deconfining phase.  However, we note that if $L< \frac{\pi}{3}R$ that chiral symmetry is again broken (unless there is some problem with an open string tachyon, to which our analysis is insensative).  We plot the curve of $\frac{L(\rh_0)}{R}$ along with the critical value $\frac{L(\rh_0^c)}{R}\approx 1.068$, and the asymptotic value of $\frac{L(\rh_0=\infty)}{R}=\frac{\pi}{3}\approx1.047$ in figure (\ref{lofr0}) to make this more clear.
One can therefore see that outside the window $\frac{\pi}{3}R < L\lessapprox 1.068 R$, chiral symmetry is broken.  When $L$ is above this window, the brane dips far enough towards the horizon that the separated brane configuration is preferred.  However, this also occurs below $\frac{\pi}{3}R$: if the asymptotic separation is below this value, there does exist a value of $\rh_0$ which allows for a connected configuration, however in this case $\rh_0$ is always less than $\rh_0^c$, and so the separated configuration is preferred.  It may be, however, that there are no solutions with $L<\frac{\pi}{3}$, and that the disconnected solution is only a local minimum.  One could argue this by starting with $L$ very large and bringing the branes together.  It is possible to tunnel from the disconnected to connected solution once $L\lessapprox 1.068 R$.  Then as one brings the ends closer together, the value of $u_0$ runs off to infinity.  In fact, from the plot in figure \ref{DeltaS}, it suggests that the effective potential is running off to $-\infty$ for smaller and smaller values of $L$.  This suggests that the solutions for $L<\frac{\pi}{3}$ are only classically stable, similar to starting with an ``inverted wine bottle'' $\phi^2-\phi^4$, and taking $\phi=0$ as a solution (when, in fact, it will tunnel out because the potential is unbounded from below).  We therefore take that there are no solutions for $L\leq \frac{\pi}{3}R$; for  $\frac{\pi}{3}R < L\lessapprox 1.068 R$, chiral symmetry is broken; and that for $L\gtrapprox 1.068R$ that chiral symmetry is broken.  This similar to the results of \cite{Aharony:2006da} where chiral symmetry was broken for  $L<0.97 R$  and restored for $L>0.97R$.

\section{Conclusions, and outlook}

The above model offers a new dual description of a model in the same universality class as large $N_c$ QCD with flavors, different from that of the SS model \cite{Witten:1998zw,Sakai:2004cn,Sakai:2005yt}.  There are certain notable similarities and differences.  First, the temperature dependence of the theory is quite striking: a thermal ``black brane'' background is only available at a specific temperature $T=\frac{1}{2\pi R}$.  Further, in the deconfined phase, the boundary conditions $\frac{L_i}{R}$ are functions only of the ratio $\frac{u_0}{u_{T}}$.  This has the interesting quality of making solutions with the same boundary conditions at infinity easy to find: if one scales $u_T$ by a factor $C$, then one must also scale $u_0$ by $C$. This has the effect that varying $u_\Lambda$ with constant boundary conditions yields no phase transitions: either chiral symmetry is always broken or restored, simply given the boundary conditions.  This allows for the possibility of making such a phase transition simultaneous.  Finally, we should also note the appearance of a finite height effective potential in the radial direction, giving a finite number of 4D mesons, unlike the case of SS.

There are also some striking similarities.  The mechanism for chiral symmetry breaking in the confined phase is identical to that of the SS model, and easy to visualize.  Further, considering the baryons as instantons in the $U(N_f)$ world volume gauge field on the D7, an almost identical mechanism for the baryon size being stabilized arises, and in both cases the size is of  the string scale.

The biggest drawback of our model is the far to simple spectrum of mesons, resulting from the finite height and flattening of all effective potentials.  One may wish to see how one could modify this setup to allow for a larger spectrum of mesons, possibly for different values of boundary conditions.  More specifically, in the confining phase, it would be interesting to see what effect turning on $P_4$ would have.  Recall that in the anti-podal embedding, the minimum of the effective potential is reached when the brane reaches all the way down to $u_\Lambda$.  Further, recall that introducing a non zero $P_4$ does not change this: the brane still reaches the minimal value $u_0=u_\Lambda$.  So, in such a situation, more effective world volume of the brane exists near where the bottom of the effective potential exists, possibly widening the bottom of the effective potential, and leading to more bound states.  Such an analysis, however, could be quite involved and need to rely on numerical methods.  A further drawback is the possible quantum mechanical instability of the system.  It would be interesting to find solutions to this as well, for example whether the backreaction of the flavor branes could help cure this problem.

Given the vast number of analysis works of the SS background, there remains many possibilities for future work.  For example, one may be curious to study the meson spectrum in the deconfining  phase similar to \cite{Peeters:2006iu}, particularly to see if one finds similar patterns to the meson melting found via quasi-normal mode analysis \cite{Hoyos:2006gb}, as well as modeling the condensation of various modes as in \cite{Aharony:2007uu}.  Also, including bare quark mass will have to follow similar lines as those in the SS model \cite{Aharony:2008an, Hashimoto:2008sr}, and studies with such deformations.  We have also not addressed any possible transition between the deconfined and confined phase.  In fact, one should imagine tuning the temperature (in the confined phase) all the way up to $T=\frac{1}{2\pi R}$.   Since this is the only temperature that admits two different solutions (with the blackening factor) this is the only possible temperature at which a phase transition happens. Note that quantum mechanically the region $T>\frac{1}{2\pi R}$ does not exist due the Hagedorn instability (neglecting the backreaction of the flavor branes).  Such a phase transition at $T=\frac{1}{2\pi R}$ would be driven simply by the presence of the flavors (as the two Euclidean actions would be identical for the gravity sector).  We would then have to subtract the two DBI actions with identical boundary conditions at infinity, but with the blackening factor on different components of the metric.  Inverting the relation between the boundary conditions $L_i$ and the $P_i$ is, however, not straightforward.  We hope to return to these questions in the future.

\section*{Acknowledgements}

We wish to thank Ofer Aharony for participating in the  early stages of this work, useful discussions and for his remarks on the manuscript.
BB also wishes to thank Gaetano Bertoldi for useful discussions.  The work of BB is supported by NSERC of Canada. The work of
J.S was supported in part by a center of excellence supported by the
Israel Science Foundation (grant number 1468/06), by a grant (DIP
H52) of the German Israel Project Cooperation, by a BSF grant  and
by the European Network MRTN-CT-2004-512194.

\end{document}